\documentclass[useAMS,usenatbib]{mn2e}

\usepackage{amsmath}
\usepackage{amssymb}
\usepackage{graphicx}
\usepackage{url}
\usepackage{multicol}
\usepackage{rotating}
\usepackage{lscape}
\usepackage{placeins}
\usepackage{float}
\usepackage{verbatim}

\usepackage{hyperref}
 \providecommand{\adsurl}[1]{\href{#1}{ADS}}

\newcommand{\lsim}{\mbox{${\,\hbox{\hbox{$ < $}\kern -0.8em \lower 1.0ex\hbox{$\sim$}}\,}$}}
\newcommand{\gsim}{\mbox{${\,\hbox{\hbox{$ > $}\kern -0.8em \lower 1.0ex\hbox{$\sim$}}\,}$}}
\newcommand{\mbi}[1]{\mbox{\boldmath$#1$}}
\newcommand{\vmax}{$V_{\rm max}$}

\begin{document}

\title[Halo mass distribution reconstruction]
{Halo mass distribution reconstruction across the cosmic web}

\author[Zhao et al.]{
  \parbox{\textwidth}{
Cheng Zhao$^1$\thanks{E-mail: zhao-c13@mails.tsinghua.edu.cn},
Francisco-Shu Kitaura$^2$\thanks{E-mail: kitaura@aip.de, Karl-Schwarzschild fellow},
Chia-Hsun Chuang$^3$\thanks{E-mail: chia-hsun.chuang@uam.es, MultiDark fellow},
Francisco Prada$^{3,4,5}$,
Gustavo Yepes$^{6}$,
Charling Tao$^{1,7}$
}
  \vspace*{4pt} \\
$^{1}$Tsinghua Center of Astrophysics \& Department of Physics, Tsinghua University, Beijing 100084, China.\\
$^{2}$Leibniz-Institut f\"{u}r Astrophysik Potsdam (AIP), An der Sternwarte 16, D-14482 Potsdam, Germany\\
$^{3}$Instituto de F\'{i}sica Te\'{o}rica, (UAM/CSIC), Universidad Aut\'{o}noma de Madrid, Cantoblanco, E-28049 Madrid, Spain\\
$^{4}$Campus of International Excellence UAM+CSIC, Cantoblanco, E-28049 Madrid, Spain\\
$^{5}$Instituto de Astrof\'{i}sica de Andaluc\'{i}a (CSIC), Glorieta de la Astronom\'{i}a, E-18080 Granada, Spain\\
$^{6}$Departamento de F\'{i}sica Te\'{o}rica, Universidad Aut\'{o}noma de Madrid, Cantoblanco, 28049, Madrid, Spain\\
$^{7}$CPPM, Universit\'{e} Aix-Marseille, CNRS/IN2P3, Case 907, 13288 Marseille Cedex 9, France
}

\date{\today}

\maketitle

\begin{abstract}
We study the relation between halo mass and its environment from a probabilistic perspective. We find that halo mass depends not only on local dark matter density, but also on non-local quantities such as the cosmic web environment and the halo-exclusion effect.
Given these accurate relations, we have developed the \textsc{hadron}-code (Halo mAss Distribution ReconstructiON), a technique which permits us to assign halo masses to a distribution of haloes in three-dimensional space. 
This can be applied to the fast production of mock galaxy catalogues, by assigning halo masses, and reproducing accurately the bias for different mass cuts.
The resulting clustering of the halo populations agree well with that drawn from the BigMultiDark $N$-body simulation: the power spectra are within 1-$\sigma$ up to scales of $k=0.2\,h\,{\rm Mpc}^{-1}$, when using augmented Lagrangian perturbation theory based mock catalogues. Only the most massive haloes show a larger deviation. For these, we find evidence of the halo-exclusion effect. A clear improvement is achieved when assigning the highest masses to haloes with a minimum distance separation. We also compute the 2- and 3-point correlation functions, and find an excellent agreement with $N$-body results. 
Our work represents a quantitative application of the cosmic web classification. It can have further interesting applications in the multi-tracer analysis of the large-scale structure for future galaxy surveys.

\end{abstract}

\begin{keywords}
(cosmology:) large-scale structure of Universe -- galaxies: clusters: general --
 catalogues -- galaxies: statistics
\end{keywords}

\section{Introduction}

Hierarchical structure formation in the Cold Dark Matter theory predicts the production of gravitationally bound compact objects called haloes \citep[][]{white1978, fry1978}. They host the galaxies we observe in our Universe according to the standard cosmological paradigm. Nevertheless, their biased relationship with respect to the underlying dark matter distribution remains still a matter of study.
In spite of the great progress made during the last decades some questions have not been fully answered yet \citep[for a review, cf.][]{cooray2002}.
We certainly know now that bias is nonlinear and scale-dependent \citep[e.g.][]{Nuza2013}.
Recent studies have shown that non-local \citep[e.g.][]{saito2014} and stochastic contributions \citep[e.g.][]{patchyp1} are also relevant in the three-point clustering statistics. In fact, parametrized bias expressions are degenerate in the two-point clustering statistics \citep[][]{patchyBS}. 

A proper bias weighting or mass weighting can reduce the variance of the clustering measurements \citep[][]{Percival2004,Seljak2009}. These methods can be applied to the galaxy surveys (e.g.,
SDSS-III/BOSS\footnote{\url{http://www.sdss.org}} \citep[][]{BOSS, Dawson2013}), of which one can estimate the biases or masses of the galaxy
sample.
The next generation of galaxy surveys (e.g. SDSS-IV/eBOSS\footnote{\url{http://www.sdss3.org/future/}}, DESI\footnote{\url{http://desi.lbl.gov/}} \citep[][]{bigboss2011}, DES\footnote{\url{http://www.darkenergysurvey.org}} \citep[][]{des2013}, LSST\footnote{\url{http://www.lsst.org/lsst/}} \citep[][]{lsst2012},
J-PAS\footnote{\url{http://j-pas.org/}} \citep[][]{jpas2014}, 4MOST\footnote{\url{http://www.aip.de/en/research/research-area-ea/research-groups-and-projects/4most}} \citep[][]{4most} or Euclid\footnote{\url{http://www.euclid-ec.org}} \citep[][]{euclid2009}) will exploit the multi-tracer analysis to constrain dark energy, the growth rate of the Universe, and hence gravity models \citep[e.g.][]{Mcdonald09b, Blake2013}.
In such a multi-tracer approach, different population of tracers of the cosmic density field are treated as independent measurements, which weighted by their distinct bias \citep[cf.][]{Mcdonald09b, Seljak2009, Hamaus2010}, will yield much tighter cosmological constraints \citep[cf. also][]{Abramo2013}.
In this context, it is fundamental to have a deep understanding of the bias for different population of tracers.

 We aim at answering several questions in this study, such as: how are haloes distributed in the cosmic web, and which properties determine the bias of different halo populations?
In particular, we investigate in this work the relation between halo mass and environment. As a practical application, we want to understand how to statistically assign halo mass to a mixed population of haloes. We will present in a subsequent publication how to extend this work to galaxy stellar masses (Kitaura et al., in prep.; Rodr\'{i}guez-Torres et al., in prep.).

The technique presented in this work has a direct application to the fast generation of mock galaxy or halo catalogues, and could be applied for the reconstruction of halo masses and density field \citep[cf. e.g.][]{Wang2009, Munoz2011, Wang2012, Munoz2012, Kitaura2013}, and to the multi-tracer analysis from galaxy redshift surveys.

This paper is structured as follows: In section \ref{sec:theo}, we first present the theoretical approach. Then, we show in section \ref{sec:num} our numerical experiments based on the BigMultiDark (\textsc{BigMD}) $N$-body simulations\footnote{\url{http://www.multidark.org/MultiDark/}} \citep[][]{Klypin2014}, which we described in section \ref{sec:ref}, followed by  the application to the generation of mock galaxy/halo catalogues based on perturbation theory in section \ref{sec:app}. Finally, we present our conclusions in section \ref{sec:con}.

\section{Theoretical approach}
\label{sec:theo}

The aim of this study is to examine the properties of the large-scale structures which statistically determine the mass of haloes.
The starting point is given by the mass function, which predicts the number of compact objects (haloes) of a certain mass (cf. pioneering works of \citet[][]{Press1974, BBKS1986}, and the later seminal works by \citet[][]{Mo1996, Sheth2004b}).
The question is which additional quantities $\{q\}$ have a significant impact on the mass of haloes from a statistical perspective. In particular, we want to answer: what determines the conditional probability distribution function of the halo mass $M_{\rm h}^i$ of an object at position $\mbi{r}_{\rm h}^i$, given  a distribution of haloes in three-dimensional space $\{\mbi{r}_{\rm h}\}$, and at redshift $z$ with cosmological parameters $\{p_{\rm c}\}$, i.e.
\begin{equation}
M_{\rm h}^i\curvearrowleft P(M_{\rm h}(\mbi{r}_{\rm h}^i)|\{\mbi{r}_{\rm h}\},\{q\},\{p_{\rm c}\},z)\,.
\end{equation}

The accuracy of this  mass assignment will have an impact on various statistical measures, such as the two- and the three-point clustering statistics. Hence, it  controls  the bias, which is the ultimate goal of this work.
Here, we follow a hierarchical approach in which we include increasingly more information in the conditional probability distribution function, evaluating at each stage the precision of the resulting bias.
From theoretical considerations based on the literature, we need to examine the impact of nonlinear local, non-local, and stochastic components of the bias \citep[e.g.][]{Press1974, Peacock1985, BBKS1986, fry1993, Mo1996, Pen1998, Dekel1999, Sheth1999, Seljak2000, Berlind2002, Smith2007, Mcdonald09, Desjacques2010, Beltran2011, Valageas2011, Elia2012, Chan2012, Baldauf2012, Baldauf2013, Angulo2014, Baldauf2014}:

\begin{enumerate}
\item We start with the simplest assumption based solely on the mass function neglecting any other information:\\ $P(M_{\rm h}^i|\{\mbi{r}_{\rm h}\},\{p_{\rm c}\},z)$.

\item The peak background split  picture models the formation of haloes of different masses in  density peaks above corresponding density thresholds \citep[][]{Kaiser1984, BBKS1986, Sheth2004a}. This theory indicates that we should consider as a next order approximation the dependence on the local density field $\rho_{\rm M}$:\\  $P(M_{\rm h}^i|\{\mbi{r}_{\rm h}\},\rho_{\rm M},\{p_{\rm c}\},z)$.

\item Recent studies have shown that non-local effects are relevant in the three-point clustering statistics \citep[e.g.][]{saito2014}. In particular, we are interested in investigating the importance of the cosmic environment in which different haloes reside. We will study this through the eigenvalues of the tidal field tensor $T$, which is a non-local measure:\\  $P(M_{\rm h}^i|\{\mbi{r}_{\rm h}\},\rho_{\rm M},T,\{p_{\rm c}\},z)$.

\item Finally,  we aim at investigating stochastic biasing \citep[e.g.][]{patchyp1,patchyBS}. This component encodes in an effective way the non-local bias contributions (in this case beyond the tidal field tensor). We will consider in particular the deviation from Poissonity. A larger dispersion than Poisson corresponds to over-dispersion, a smaller one to under-dispersion, which are due to the positive or negative correlation on sub-grid scales, respectively \citep[e.g.][]{Peebles1980, Somerville2001, CasasMiranda2002, Baldauf2013, patchyp1}. We will focus in this study on the minimum separation between haloes $\Delta r_{\rm min}^{M}$: \\
$P(M_{\rm h}^i|\{\mbi{r}_{\rm h}\},\rho_{\rm M},T,\Delta r_{\rm min}^{M},\{p_{\rm c}\},z)$.
\end{enumerate}

In the next section we will investigate the relevance of the different bias components $\{q\}=\{\rho_{\rm M},T,\Delta r_{\rm min}^{M}\}$, based on the analysis of a large $N$-body cosmological simulation.

\section{Reference $N$-body simulation and halo catalogues}
\label{sec:ref}

In particular, we employ the dark matter particle and halo catalogues at redshift $z = 0.5618$, extracted from one of the \textsc{BigMD} simulations, which was performed using the TreePM $N$-body code \textsc{gadget-2} \citep[][]{gadget} with $3840^3$ particles in a volume of $(2.5\, h^{-1}\mathrm{Gpc})^3$, within the framework of Planck $\Lambda$CDM cosmology with $\{\Omega_m = 0.307115, \Omega_b =0.048206, \sigma_8 = 0.8288, n_s = 0.96 \}$, and the Hubble parameter ($H_0 \equiv 100\,h\,\mathrm{km}\, \mathrm{s}^{-1} \mathrm{Mpc}^{-1}$) given by $h=0.6777$.

We have two sets of halo catalogues constructed by using two different halo finders, the spherical overdensity Bound Density Maxima (BDM) \citep[][]{Klypin1997, Gottloeber1998, Riebe2011} and the Friends-of-Friends (FoF) \citep[][]{Gottloeber1998, Riebe2011} halo finders with linking length $l=0.17$  times the mean interparticle distance (0.11 $h^{-1}$Mpc). We select the BDM haloes and subhaloes by \vmax{} (i.e., maximum circular velocity), and use mass for FoF haloes, to construct complete samples from both halo catalogues with number density $3.5 \times 10^{-4}\,h^3\,{\rm Mpc}^{-3}$, as that for typical Luminous Red Galaxies in large-scale surveys.

We follow \citet[][]{Klypin2011} for the definition of \vmax{}:
\begin{equation}
V_{\rm max} = \left. \sqrt{\frac{G M(<r)}{r}}\right|_{\rm max} .
\end{equation}
We choose this quantity as a proxy for halo mass, since \cite[cf.][]{Prada2012}, 1) it is a more reliable quantity than the mass defined at a given over-density, and
2) it is better for the characterization of haloes when relating them to the galaxies inside. It has a more direct relation with observational quantities, such as luminosity or stellar mass, that are used for defining galaxy catalogues with Halo Abundance Matching (HAM) modelling \citep[][]{Trujillo-Gomez2011}.
We will focus in this work on BDM (sub)haloes and show some results using FoF haloes in Appendix \ref{sec:fof}.

\section{The halo mass environmental dependence}
\label{sec:num}

Haloes are generally identified in an $N$-body dark matter density field using the so-called halo-finder algorithms. This is essentially an estimate of the halo bias, which encodes a certain relation between halo masses and the dark matter density field. 

Therefore, let us start by studying the bias from $N$-body cosmological simulations that provide both dark matter particles and the corresponding halo catalogues. To follow the analysis proposed in the previous section we will investigate the clustering statistics of different populations of haloes conditioned on different degrees of information.

To evaluate the accuracy of the  bias, we will start with the two-point statistics in Fourier space.
 In particular, we compute power spectra of different populations of haloes, whose amplitudes are essentially a direct estimate of the bias factor \citep[e.g.][]{Cen1992}, and their shapes show the scale-dependency. In this work, we adopt the cloud-in-cells particle assignment scheme (CIC) for haloes with grid size of $512^3$ to perform the fast Fourier transform, and then compute the power spectra with aliasing and shot noise corrections taken into account \citep[cf.][]{Jing2005}.

Our aim is to reduce systematic deviations on the power spectra to a few percent on scales relevant to baryon acoustic oscillations, i.e. $k\lsim0.2\,h$\,Mpc$^{-1}$. We study in this section different mass assignment procedures, and show how this goal can be achieved, provided we take into account density--mass relation, cosmic web environment, and halo-exclusion.
Let us now perform our numerical analysis trying to recover the masses of a given three-dimensional distribution of haloes going through the steps outlined in \S \ref{sec:theo}.

\begin{figure}
\centering
\includegraphics[width=.47\textwidth]{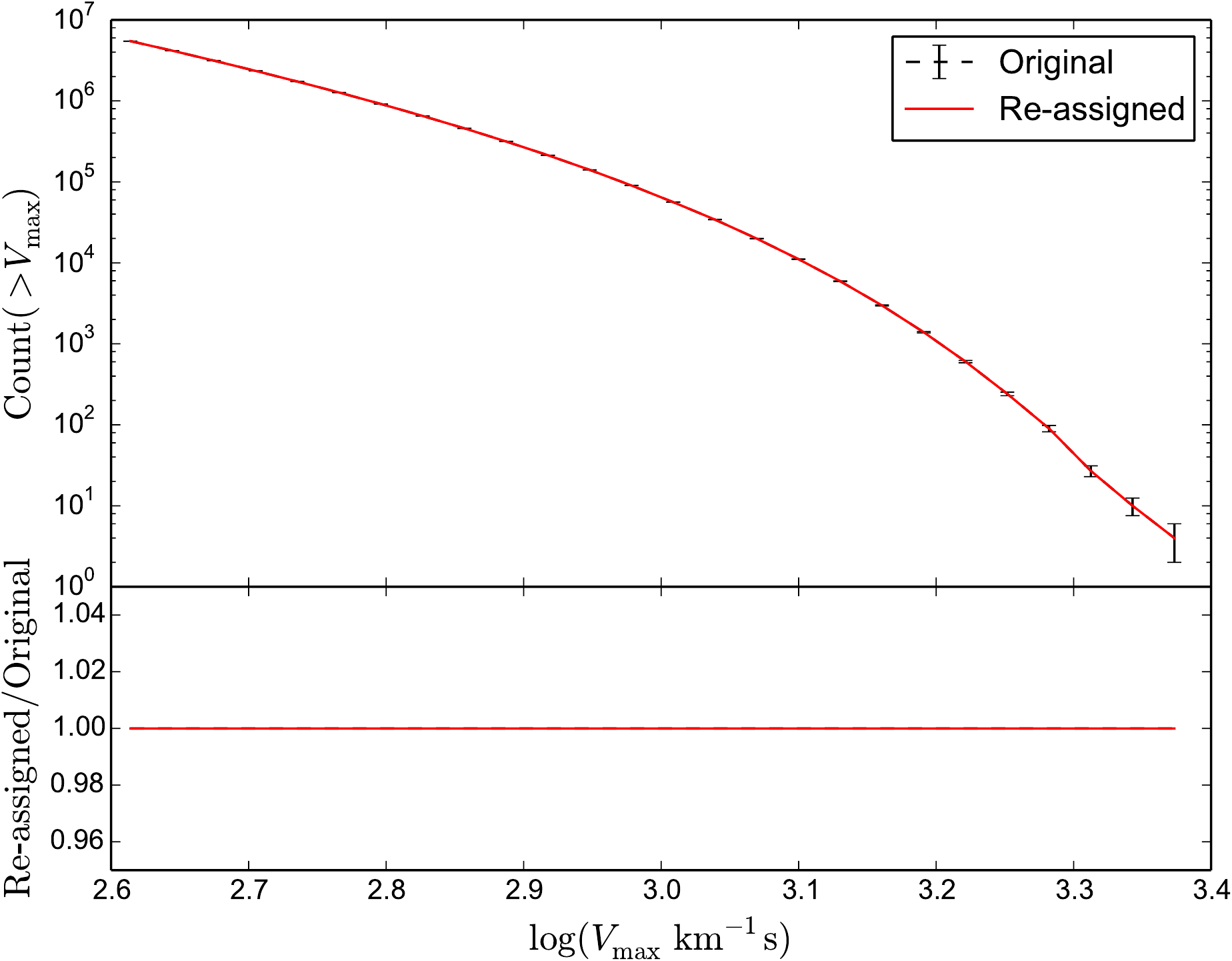}
\caption{Halo mass (\vmax{}) functions of the original data drawn from the \textsc{BigMD} simulation and re-assigned halo catalogue. They agree with each other by construction. The error bars show Poisson errors.}
\label{fig:mf}
\end{figure}

\begin{figure}
\centering
\includegraphics[width=.47\textwidth]{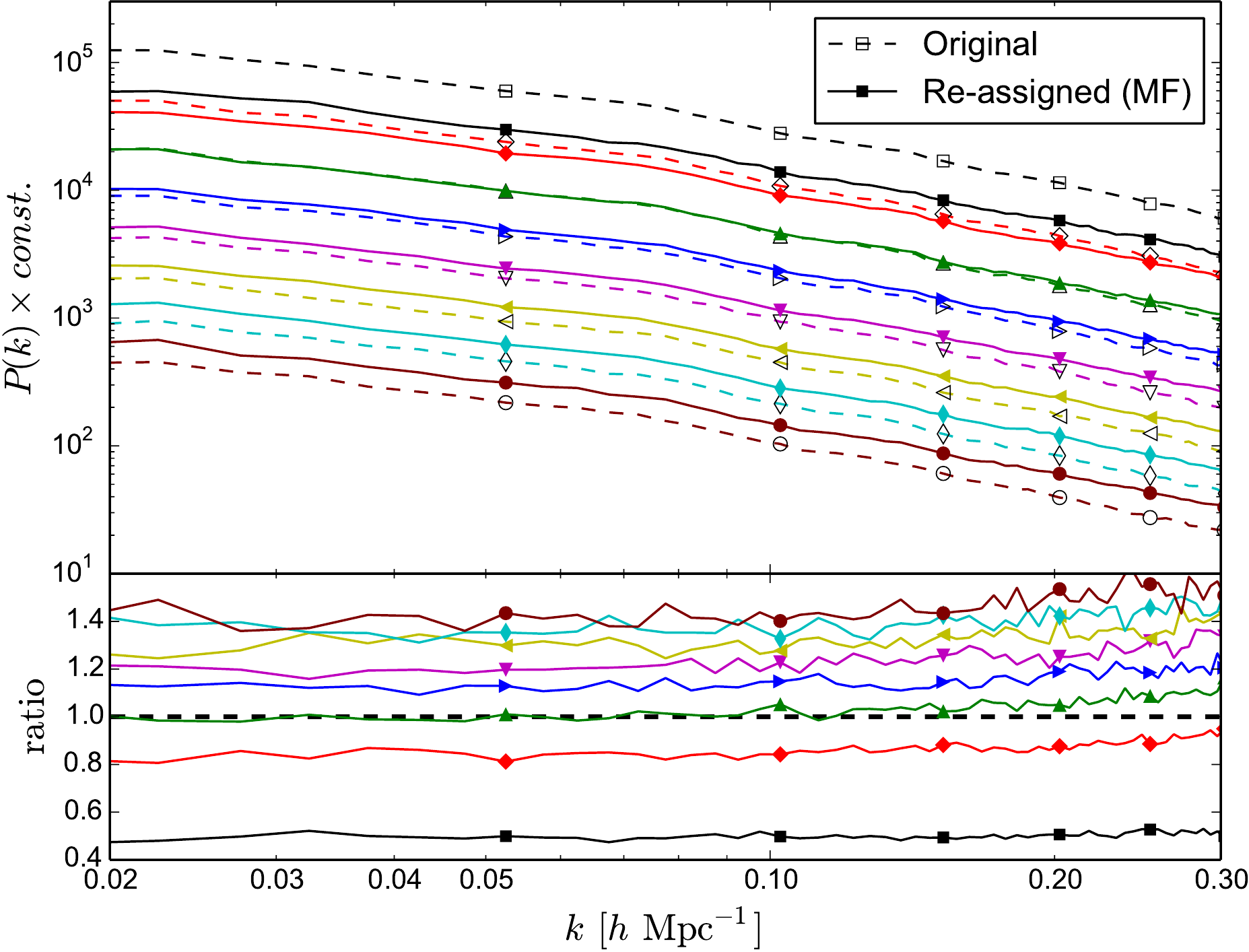}
\caption{Power spectra of the halo sub-samples discussed in the text, with different halo \vmax{}. Dashed lines indicate the power spectra for the sub-samples drawn from the \textsc{BigMD} BDM catalogue, while solid lines correspond to that after applying the mass assignment procedure. MF stands for the mass (\vmax{}) function used in this case, see Fig.~\ref{fig:mf}. The different colour codes correspond to the different \vmax{} bins of the sub-samples increasing from the bottom to the top lines (\{[$<$410), [410-427), [427-447), [447-472), [472-501), [501-550), [550-640), [$\geq$640]\}\,km\,s$^{-1}$). For visualisation purposes the power spectra corresponding to different \vmax{} bins have been multiplied with different constants in the upper panel to enlarge the differences}.
\label{fig:pk_mf}
\end{figure}

\begin{figure*}
\centering
\includegraphics[width=.9\textwidth]{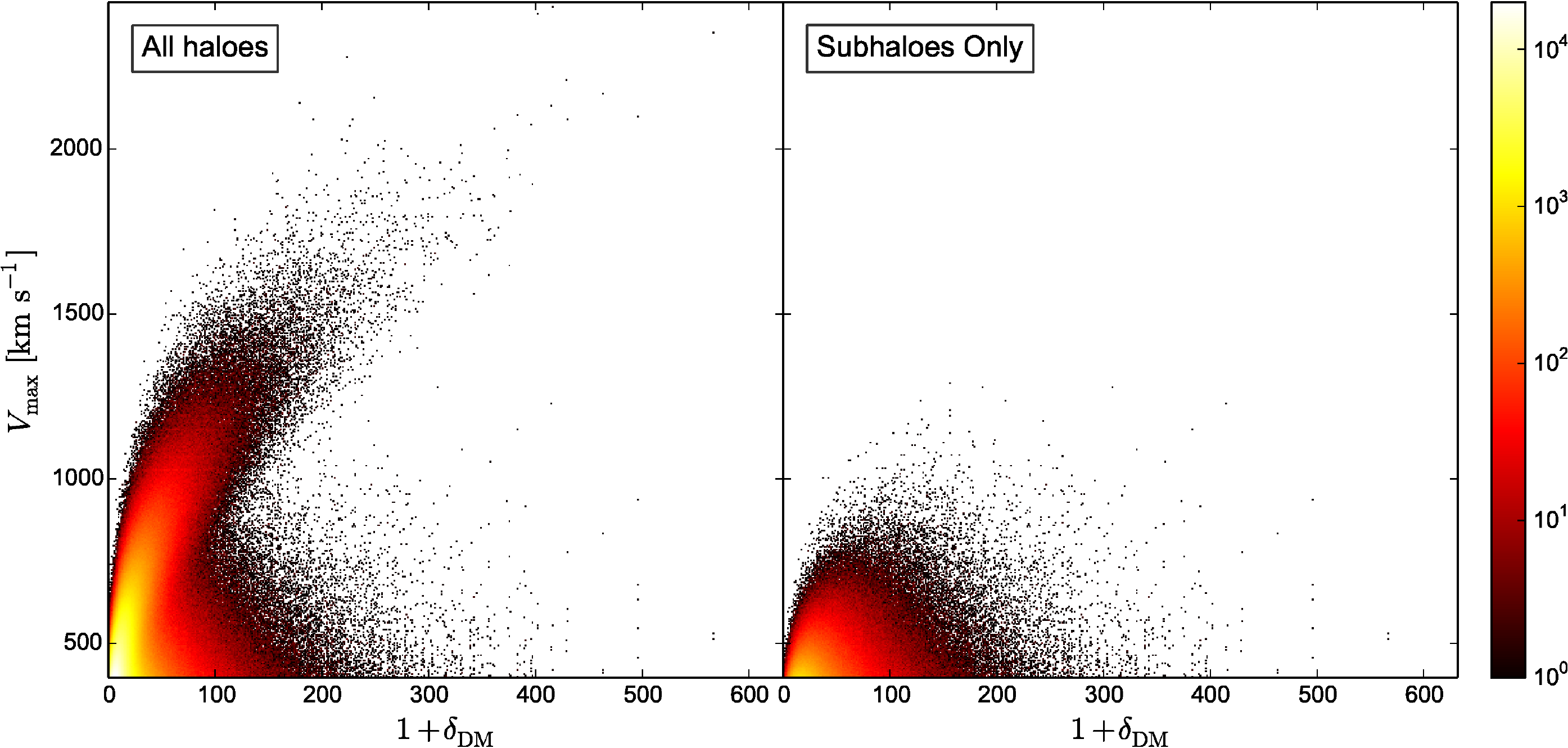}
\caption{Number of (sub)haloes in the \textsc{BigMD} simulation with certain \vmax{} and local DM density.}
\label{fig:vd}
\end{figure*}

\subsection{Mass function}
\label{sec:mf}

We start by studying the halo \vmax{} reconstruction considering only the mass (\vmax{}) function, i.e. the conditional probability function $P(M_{\rm h}^i|\{\mbi{r}_{\rm h}\},\{p_{\rm c}\},z)$. In particular, each halo gets a \vmax{} assigned following the function shown in Fig.~\ref{fig:mf} regardless of its location.   
In this case, the \vmax{} cumulative function is reproduced by construction.

To verify the performance of the halo \vmax{} reconstruction, we divide both the original BDM halo catalogue and the re-assigned one, into 8 sub-samples of halo \vmax{} and compare the clustering for each sub-sample. In particular, we cut the catalogues in \vmax{} bins of \{[$<$410), [410-427), [427-447), [447-472), [472-501), [501-550), [550-640), [$\geq$640]\} km\,s$^{-1}$ to have similar number of haloes in each bin.
We can see in Fig.~\ref{fig:pk_mf} how the power spectra of the different sub-samples have amplitudes which disagree with the true ones.  The degree of the systematic deviation depends on the mass of the halo population. Low-mass haloes (\vmax{}$\sim500$ km\,s$^{-1}$) yield an overestimation of the bias (the ratio of the reconstructed  and the true power spectrum is greater than one). Only haloes in the range around  \vmax{}$\sim500-550$ km\,s$^{-1}$ are closely unbiased, and more massive haloes lead to an underestimation of the bias  (the ratio of the reconstructed  and the true power spectrum is lower than one).
Nevertheless, the deviations of up to $\sim$40\% throughout the full $k$-range in the power spectra hint that we need to consider additional environmental indicators to make a precise mass assignment. 

\subsection{Density--halo mass relation}
\label{sec:vmax1}

Let us now investigate the relation between the halo mass (or equivalently \vmax{}) and the local dark matter density.
We employ the CIC scheme for dark matter particles in the \textsc{BigMD} simulation with a grid size of $960^3$ to obtain the dark matter density field, and then distribute haloes to the same mesh using the nearest-grid-point scheme (NGP), since as our application, the mocks employ CIC while computing the  dark matter density field, and we need to use integers for the mass assignment procedure. For each halo, the local dark matter density is defined by the number density contrast of dark matter particles in the corresponding cell, i.e., $1 + \delta_{\rm DM} = \rho / \bar{\rho}$, with $\delta_{\rm DM}$ being the density fluctuations, $\rho$ the density, and $\bar{\rho}$ the mean density.

\begin{figure}
\centering
\includegraphics[width=.47\textwidth]{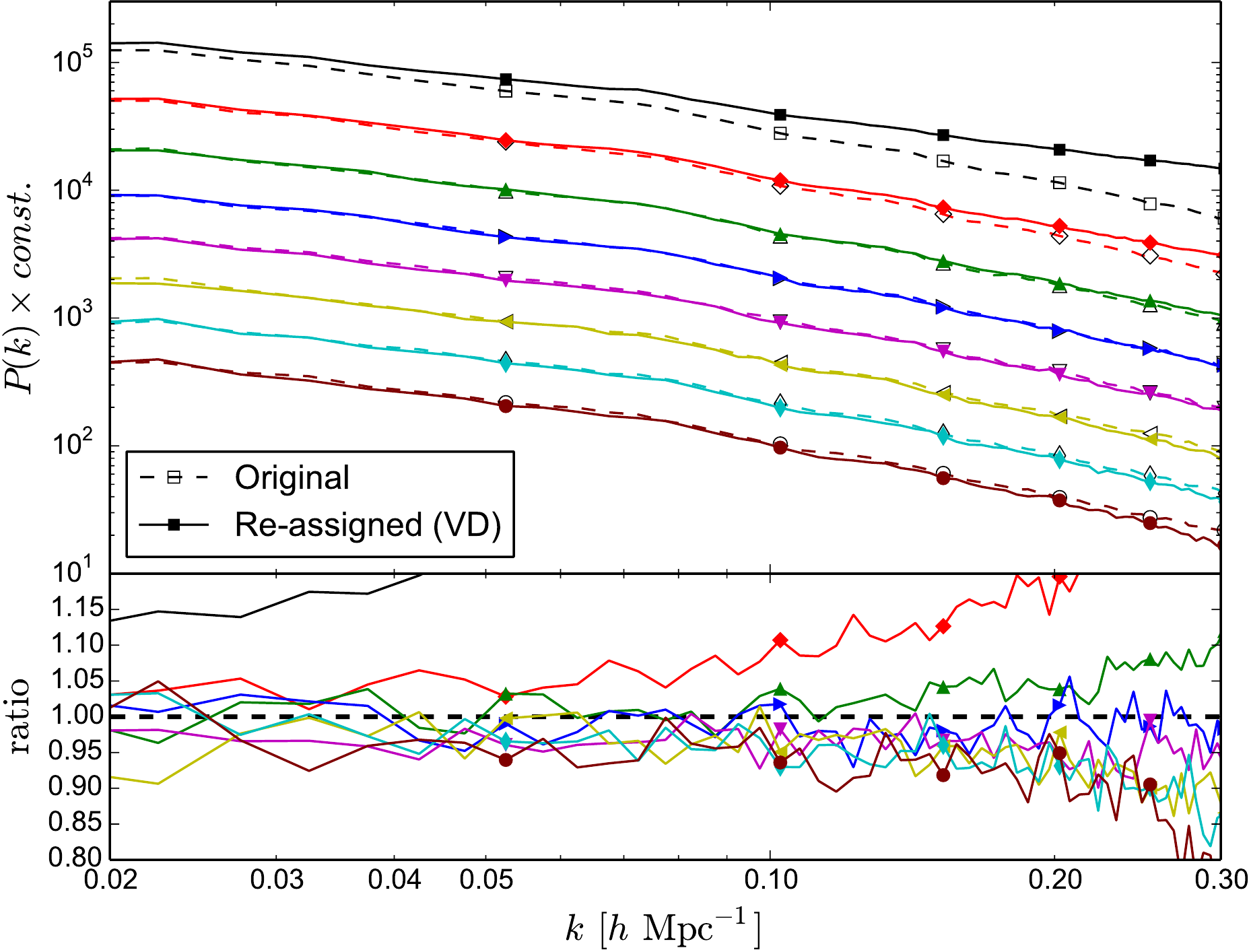}
\caption{Power spectra of the sub-samples with different halo \vmax{}. Same convention as in Fig.~\ref{fig:pk_mf}. VD stands for the \vmax{}--density relation used in this case.}
\label{fig:pkv1}
\end{figure}

The left panel in Fig.~\ref{fig:vd} shows the distribution of all haloes, and subhaloes only in the \vmax{} vs. local dark matter density plane, where we adopt 500 bins for both \vmax{} and dark matter density. Massive haloes tend to be located in dense environments, obeying a power law. Interestingly, there is a halo mass range suppressed at high densities, indicating that these moderately massive haloes are merged to more massive ones in such environments. Another remarkable feature is that there are two branches in the high density regions, indicating two different groups of haloes residing in the same dark matter environment.
The haloes in the BDM catalogue can be divided into distinct and subhaloes. Thus, we find that haloes in the low-\vmax{} region are predominantly sub-structures of those with higher \vmax{} (see right panel of Fig.~\ref{fig:vd}). This is why the haloes in very dense regions (e.g. $1+\delta_{\rm DM} > 400$) are located in discrete bins of constant $1 + \delta_{\rm DM}$ in Fig.~\ref{fig:vd}, as can be more clearly seen in the outliers at high $1 + \delta_{\rm DM}$ values. 
This relation shows the probability of finding a halo with a given \vmax{} in a certain dark matter density environment.

We then keep the local dark matter density of each halo, and assign \vmax{} to the haloes according to the extracted probability, ignoring their original \vmax{}. The new catalogue has the same \vmax{}--density relation as the original one, given the same \vmax{} and density bins. The re-assignment procedure is equivalent to a shuffling of the halo \vmax{} in each dark matter density bin.

A clear improvement is found with respect to the previous results given in \S \ref{sec:mf}, as can be seen in the power spectra shown in Fig.~\ref{fig:pkv1}. Nevertheless,  the lowest and the next to largest \vmax{} bins still  show systematic deviations of about 10 and 15\% up to $k\sim0.15\,h$\,Mpc$^{-1}$, respectively (and increasingly larger towards larger $k$). The largest \vmax{} bin shows even a deviation exceeding 30\% at $k\sim0.15\,h$\,Mpc$^{-1}$. This indicates that we still need to include more information to reach the desired accuracy of $\sim10\%$.

\begin{figure}
\centering
\includegraphics[width=.45\textwidth]{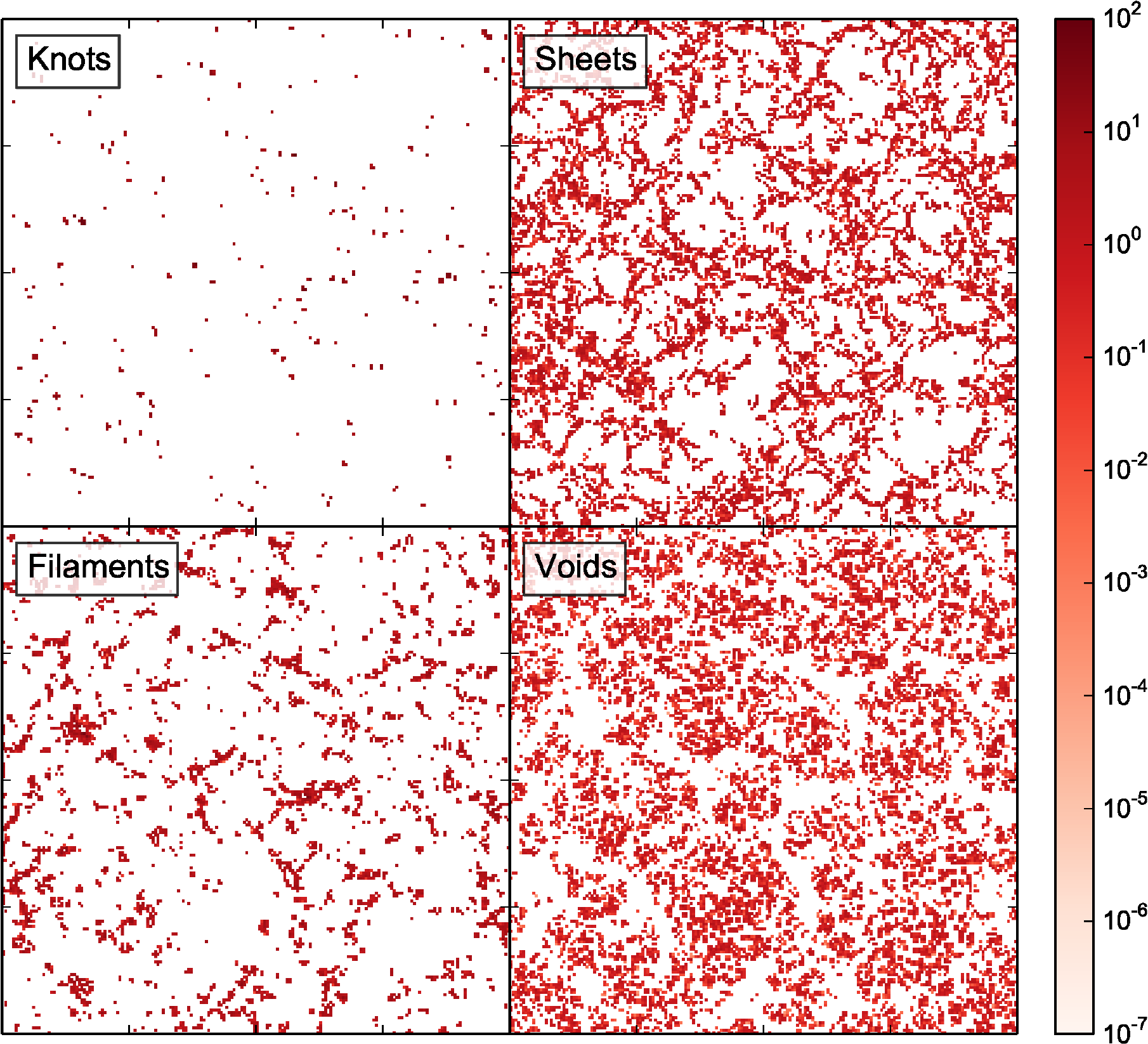}
\caption{Dark matter density field in the \textsc{BigMD} simulation corresponding to different cosmic web structures, as indicated in the legend of the panels.}
\label{fig:cw}
\end{figure}

\subsection{Cosmic web environment \label{sec:vmax2}}
\label{sec:cw}

As the next step, we now include non-local indicators. 
The tidal field tensor includes second order non-local information, and its eigenvalues have been used to make a cosmic web classification (e.g. \cite{Hahn2007,Forero-Romero2009,Aragon2010, Hoffman2012}).
Since different types of cosmic web structures can have the same local matter density, our analysis in the previous section  mixed haloes residing in different structures and thus with different biases. Therefore, we split the haloes according, not only to their local dark matter density, but also to the type of cosmic web structures they live in.

In particular, we classify cosmic web structures following \citet[][]{Hahn2007} and \citet[][]{Forero-Romero2009}. From the dark matter density field, we obtain the gravitational potential $\phi$ from the Poisson equation, and construct the tidal field tensor
\begin{equation}
T_{i j} = \frac{\partial^2 \phi}{\partial x_i \partial x_j}.
\end{equation}
If all the three eigenvalues of $T_{i j}$ are above (below) a certain threshold ($\lambda_{\rm th}$), then the local structure is collapsing (expanding) in all directions, forming a knot (void). When one (two) eigenvalue is above the threshold, we have filament-like (sheet-like) structures.

A slice of the dark matter density field for $\lambda_{\rm th} = 0$ is shown in Fig.~\ref{fig:cw}, for different types of cosmic web structures. Although voids, sheets, and filaments occupy most of the volume, the majority of the haloes are located in knots. To be more precise, we further classify knots by their total enclosed mass into several classes. To compute the mass of each knot, we adopt a simple FoF algorithm resolving single knots from the dark matter density field. The mesh cells of knots are marked during the cosmic web classification, and subsequently all marked cells that are next to each other are merged to construct a single knot. The mass of the knot is then proportional to the sum of dark matter mass of all the connected cells. Our analysis shows that, for the halo masses considered in this study, the distinction between non-knots structures does not add any information. Therefore, we combine the rest of structures into a single specie. This situation may however be different when considering lower mass haloes.

\begin{figure}
\centering
\includegraphics[width=.45\textwidth]{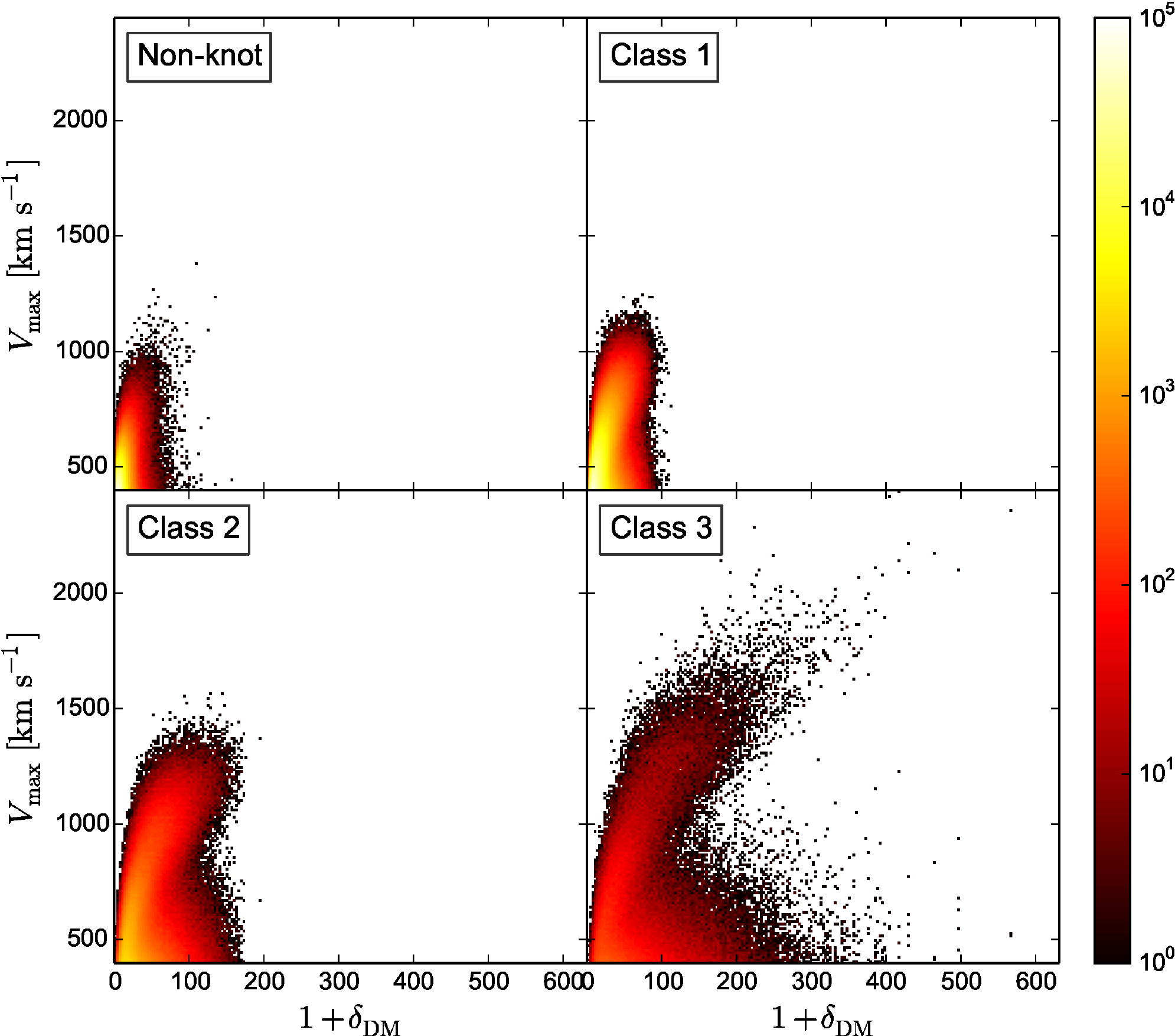}
\caption{\vmax{}--density relation of different cosmic web structure classes. The mass of knots increase from class 1 to class 3.}
\label{fig:kv}
\end{figure}

With the different types of cosmic web structures resolved from the dark matter density field, we can then extract the relation between the halo \vmax{} and the dark matter density for each class, as shown in Fig.~\ref{fig:kv}. In this plot, only three classes of knots are shown to illustrate the various \vmax{}--density relations as a function of the mass of knots. Nevertheless, the total number of classes in the re-assignment procedure exceeds 500.

\begin{figure}
\centering
\includegraphics[width=.47\textwidth]{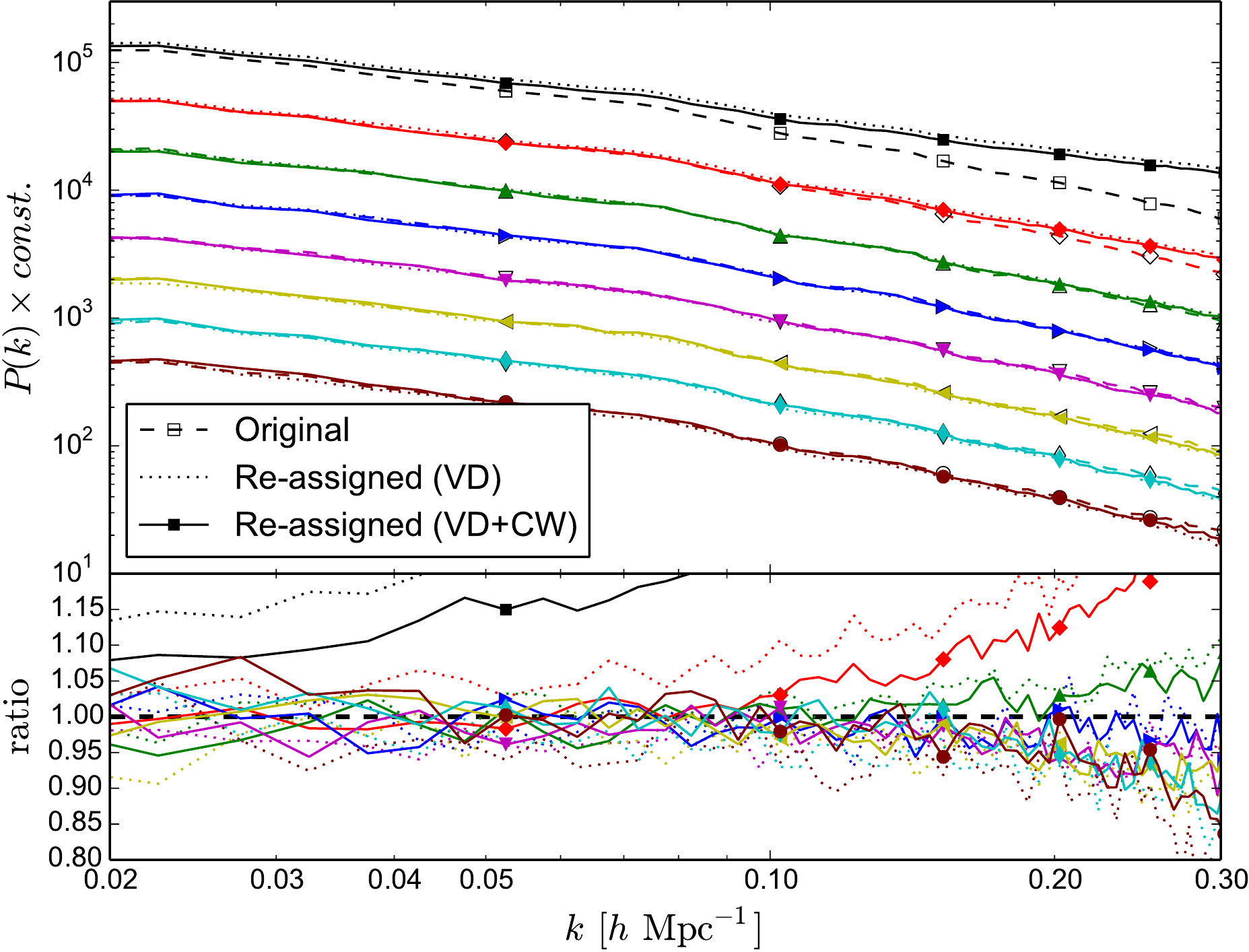}
\caption{Power spectra of the sub-samples with different halo \vmax{}.  Same convention as in Fig.~\ref{fig:pk_mf}. CW stands for the cosmic web classification additionally used in this case. The dotted lines indicate the results of \S \ref{sec:vmax1}.}
\label{fig:pkv2}
\end{figure}

We build the same number of classes for the catalogue without halo mass, and then separately assign masses to haloes in different classes. This procedure leads to a clear improvement, as can be seen in Fig.~\ref{fig:pkv2}.
The power spectra are now compatible with the true ones up to $k\sim0.15\,h$\,Mpc$^{-1}$ within 5\%.  Nevertheless, the more massive haloes with the largest \vmax{} still show an increasing systematic deviation towards high $k$. The next to the largest mass bin deviates by $\sim$10\% at  $k=0.2\,h\,{\rm Mpc}^{-1}$, and more than  20\% at $k=0.25\,h\,{\rm Mpc}^{-1}$, while the largest mass bin deviates already more than 20\%  at  $k=0.1\,h$\,Mpc$^{-1}$. 
Let us therefore focus in the next section on the most massive haloes.

\subsection{Halo-exclusion}
\label{sec:vmax3}

According to the results of the previous section, the information of the mass function, local density, and cosmic web environment is not enough to accurately determine the mass of the most massive haloes. Still, a clear systematic deviation is present in the power spectra with a tendency to overestimate the clustering of massive objects  (larger power towards high $k$). 

At this stage, we should note that the mass assignment we have conducted, although depending on the environment, followed a random procedure, hereby ignoring deviations from the Poisson distribution beyond the ones present in the actual three-dimensional distribution of haloes. In this sense, we did not distinguish between distinct haloes and subhaloes. A distinct halo, and its subhaloes sharing the same dark matter environment, could get  the same assigned mass with equal probability. Thus the re-assigned catalogue, does not prevent two massive haloes from being arbitrarily close to each other, and hence leads to a higher power spectrum. But actually, the halo-exclusion effect affecting massive haloes yields under-dispersed (dispersion smaller than Poisson) distributions \citep[e.g.][]{Somerville2001, CasasMiranda2002, Baldauf2013}.

We therefore consider now the minimum separation between massive haloes $\Delta r_{\rm min}^{M}$. In particular, to separate massive haloes in the \vmax{} assignment procedure, we perform an additional operation by setting a \vmax{} threshold. In order to distribute haloes with \vmax{} above the threshold to different cells as far as possible, we follow a top-down procedure, beginning with the highest \vmax{} and continuing towards lower values, randomly selecting un-occupied cells to ensure that there is only one such halo in each cell. However, for a relatively low \vmax{} threshold, there might not be available cells for all the haloes. In this case, we assign two or even more haloes to one cell.

\begin{figure}
\centering
\includegraphics[width=.47\textwidth]{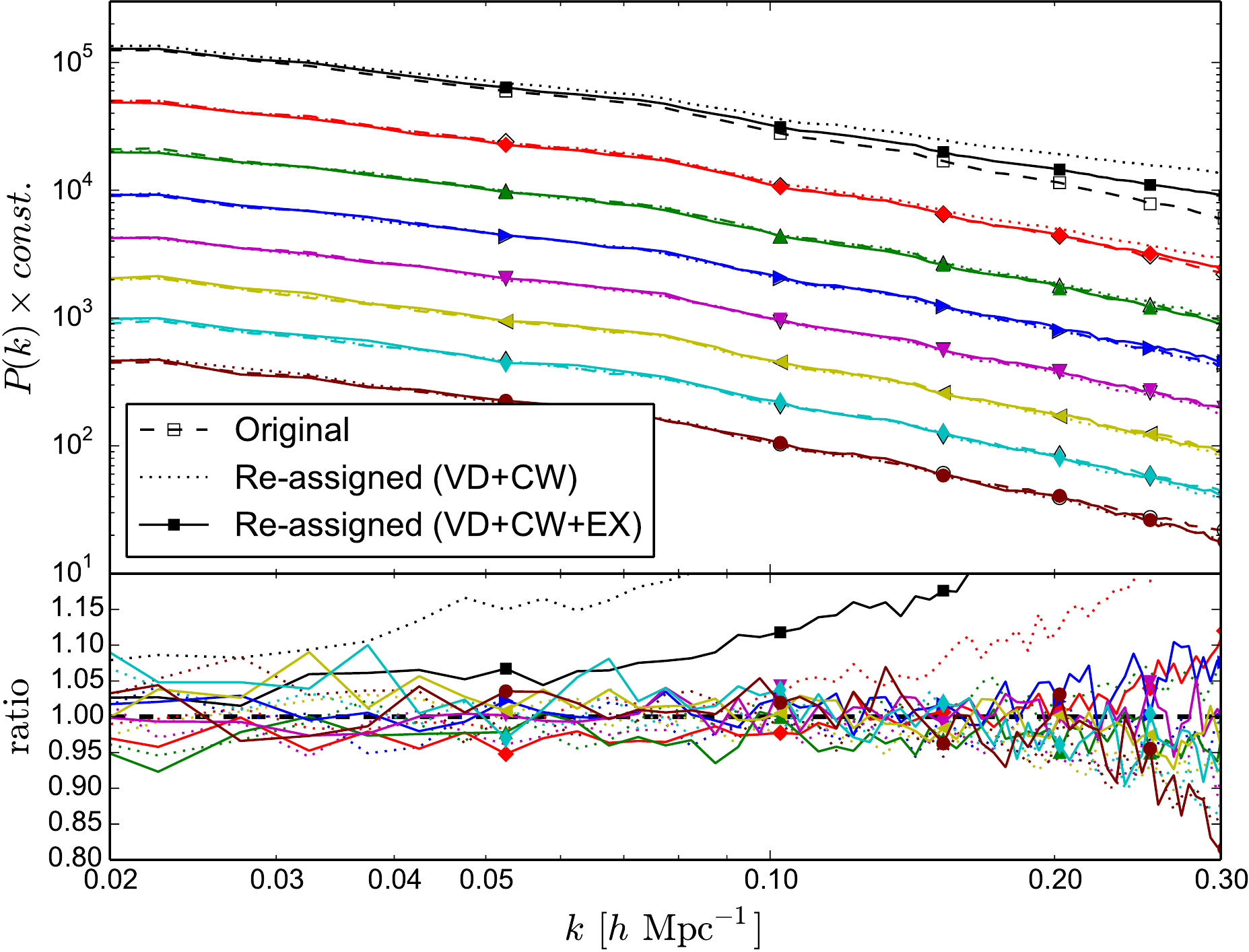}
\caption{Power spectra of the sub-samples with different halo \vmax{}.   Same convention as in Fig.~\ref{fig:pk_mf}. EX stands for the halo-exclusion additionally used in this case. The dotted lines indicate the results in \S \ref{sec:vmax2}}
\label{fig:pkv3}
\end{figure}

This procedure can be refined using the distribution of separation between haloes for different mass bins  extracted from $N$-body simulations and applied in a stochastic way, as we do with the \vmax{}--density relation. We leave such a study for future work.
Nevertheless, the procedure described above serves to test our hypothesis and leads to already clear improvements. In particular, we find that most of the mass (\vmax{}) bins show power spectra which are compatible with the true ones within about 5\%, up to $k\sim0.2\,h$\,Mpc$^{-1}$. Only the most massive bin still shows a clear systematic deviation which has been reduced to less than 15\% up to $k\sim0.15\,h$\,Mpc$^{-1}$, indicating that we still need to increase the halo-exclusion effect for this mass bin (see Fig.~\ref{fig:pkv3}).

\begin{figure}
\centering
\includegraphics[width=.47\textwidth]{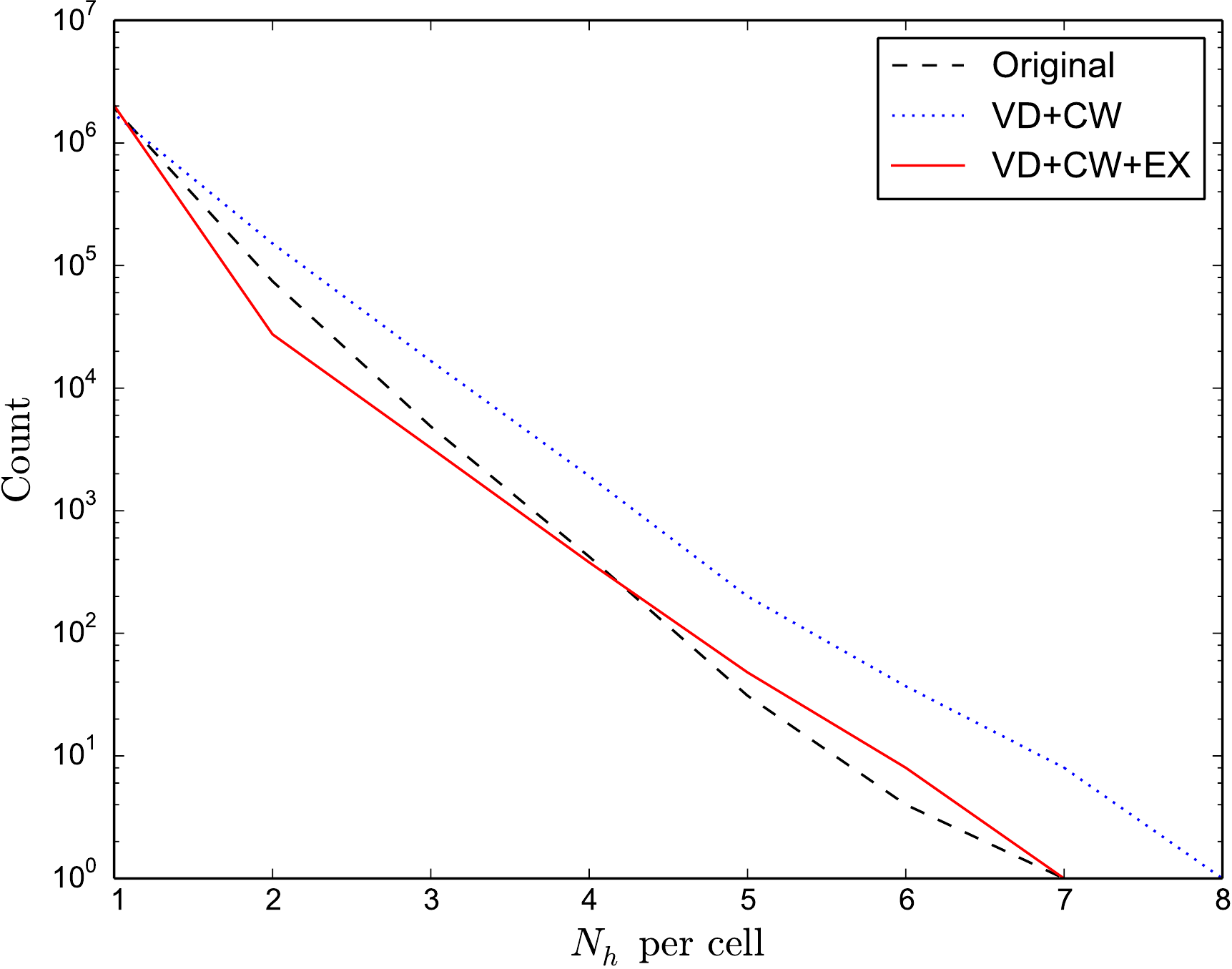}
\caption{Probability distribution function of haloes with \vmax{} above 500\,km\,s$^{-1}$.}
\label{fig:pdf_500}
\end{figure}

As a further demonstration of the halo-exclusion correction performed in this section, we investigate the one-dimensional probability distribution function (PDF) of haloes with \vmax{} above a threshold of 500\,km\,s$^{-1}$, as shown in Fig.~\ref{fig:pdf_500}. Here we can see that the PDF matches the true one only after applying the correction. We know from previous studies that an accurate PDF is especially important regarding the higher order statistics \citep[e.g.][]{patchyBS}.
An analogous analysis has been conducted with FoF haloes and the results are shown in Appendix \ref{sec:nbody}.

 Let us now have a look at applications of this method and its performance in terms of 2- and 3-point statistics in the next section, computed with the \textsc{ntropy-npoint} software, which is an exact n-point calculator using a kd-tree framework with true parallel capability and enhanced routine performance \citep[][]{Gardner2007, McBride2011}.

\section{Application: mock catalogues}
\label{sec:app}

In the previous section, we have developed a prescription which allows us to accurately (within 15\% up to $k\sim0.15\,h$\,Mpc$^{-1}$) describe the complex halo mass--environment dependence. We dub this method the Halo mAss Distribution ReconstructiON code (\textsc{hadron}).
As an application of our method, we study in this section the assignment of masses to mock halo catalogues constructed with perturbation theory. In particular, we consider two methods, the PerturbAtion Theory Catalogue generator of Halo and galaxY distributions (\textsc{patchy}) \citep[][]{patchyp1,patchyBS} and the Effective Zel'dovich approximation mocks (\textsc{EZmocks}) \citep[][]{ezmockp}. Both methods provide mock halo catalogues calibrated with the \textsc{BigMD} BDM and FoF halo catalogues, as well as the dark matter particle distributions.
While \textsc{patchy} includes an explicit Eulerian nonlinear and stochastic bias description,  \textsc{EZmocks} uses effective modifications of the initial conditions and bias modelling to reproduce the bias of objects in the final catalogue. 
The dark matter density field used in \textsc{EZmocks} is given by the Zel'dovich approximation \citep[][]{ZA}, while \textsc{patchy} uses ALPT \citep[][]{ALPT}.
The different approximations have an impact on the accuracy of the bias, as we will show below. 

\subsection{The HADRON-code}
\label{sec:hadron}

We outline below the steps included in the \textsc{hadron}-code to assign masses to haloes constructed with approximate gravity solver based mock generators.

\begin{enumerate}
\item \label{algo_1}
First, we compute the density field and cosmic web structures (knots, sheets, filaments, and voids) according to the dark matter particles from a reference $N$-body simulation. Then, we further classify the knots into different classes according to their enclosed mass (see details in \S \ref{sec:cw}). As a result, we obtain the density ($\rho_{\rm DM}$) and cosmic web classification type ($t_{\rm CW}$) for each cell.

\item \label{algo_2}
Second, we compute the number of haloes in each density and cosmic web classification type bin according to the halo catalogue from the reference $N$-body simulation.

\item \label{algo_3}
Third, we take the dark matter particles according to the approximate catalogue mock generator, and compute  the density and cosmic web type in an analogous way to step (i). Since these quantities are different for simulations and mocks, we rank order those from mocks, in order to have an equivalent population of haloes in each $\rho_{\rm DM}$ and $t_{\rm CW}$ bin.

\item \label{algo_4}
Fourth, we apply halo-exclusion to the halo catalogue from the mock generator, i.e. we assign mass above the threshold to haloes (see details in \S \ref{sec:vmax3}).

\item \label{algo_5}
Finally, we assign the mass to the rest of the haloes. For each mock halo without mass (some of them have already acquired a mass in step~\ref{algo_4}), we find the local density $\rho_{\rm DM}$ and cosmic web classification type $t_{\rm CW}$. From step~\ref{algo_2} we have the distribution of halo masses for a given $\rho_{\rm DM}$ and $t_{\rm CW}$, we then choose the mass of this halo with the probability from the corresponding distribution.
\end{enumerate}

\begin{figure}
\centering
\includegraphics[width=.47\textwidth]{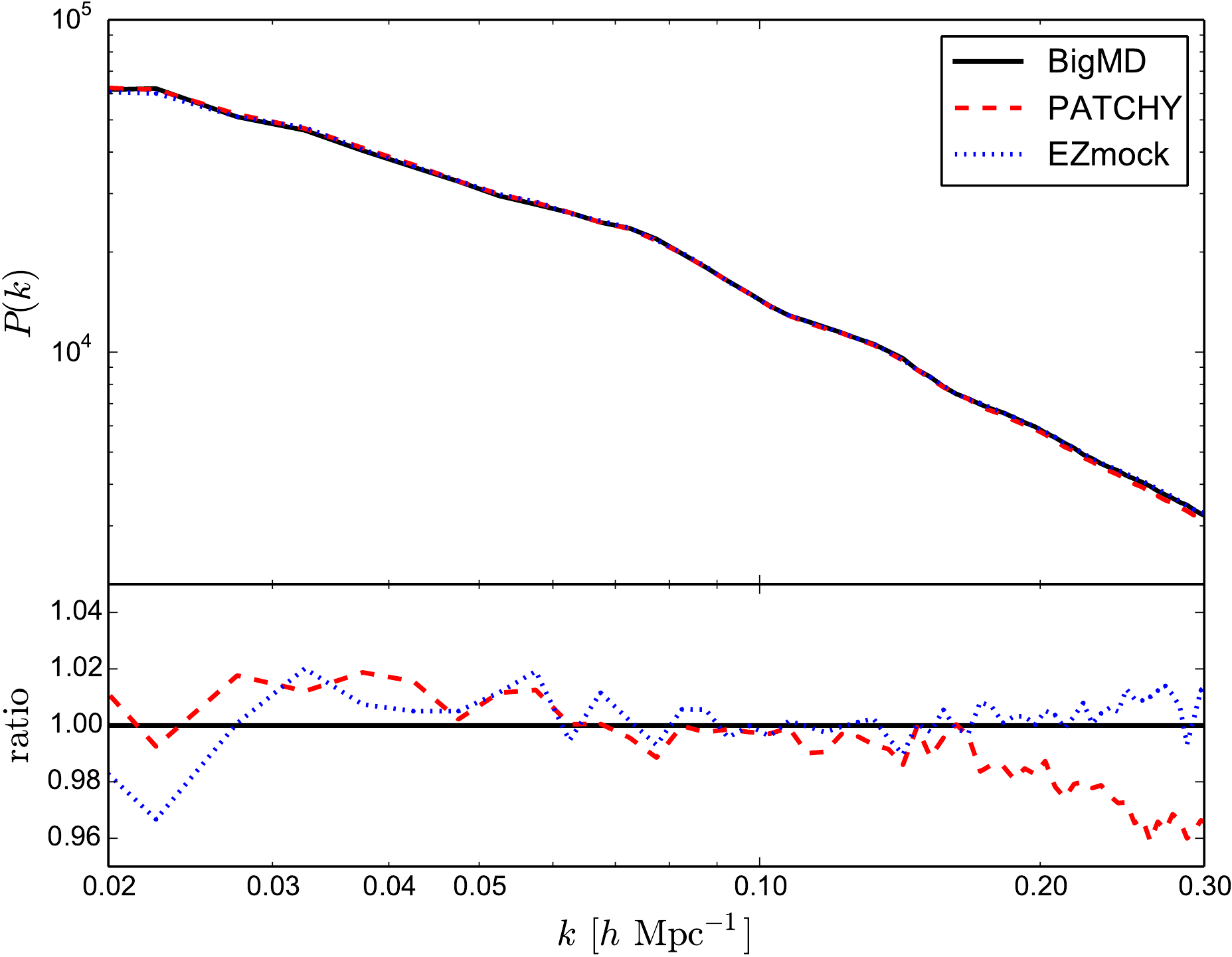}
\caption{Comparison of power spectra for the mock catalogues before \vmax{} assignment.}
\label{fig:bdm_all}
\end{figure}

\begin{figure*}
\begin{tabular}{cc}
\includegraphics[width=.47\textwidth]{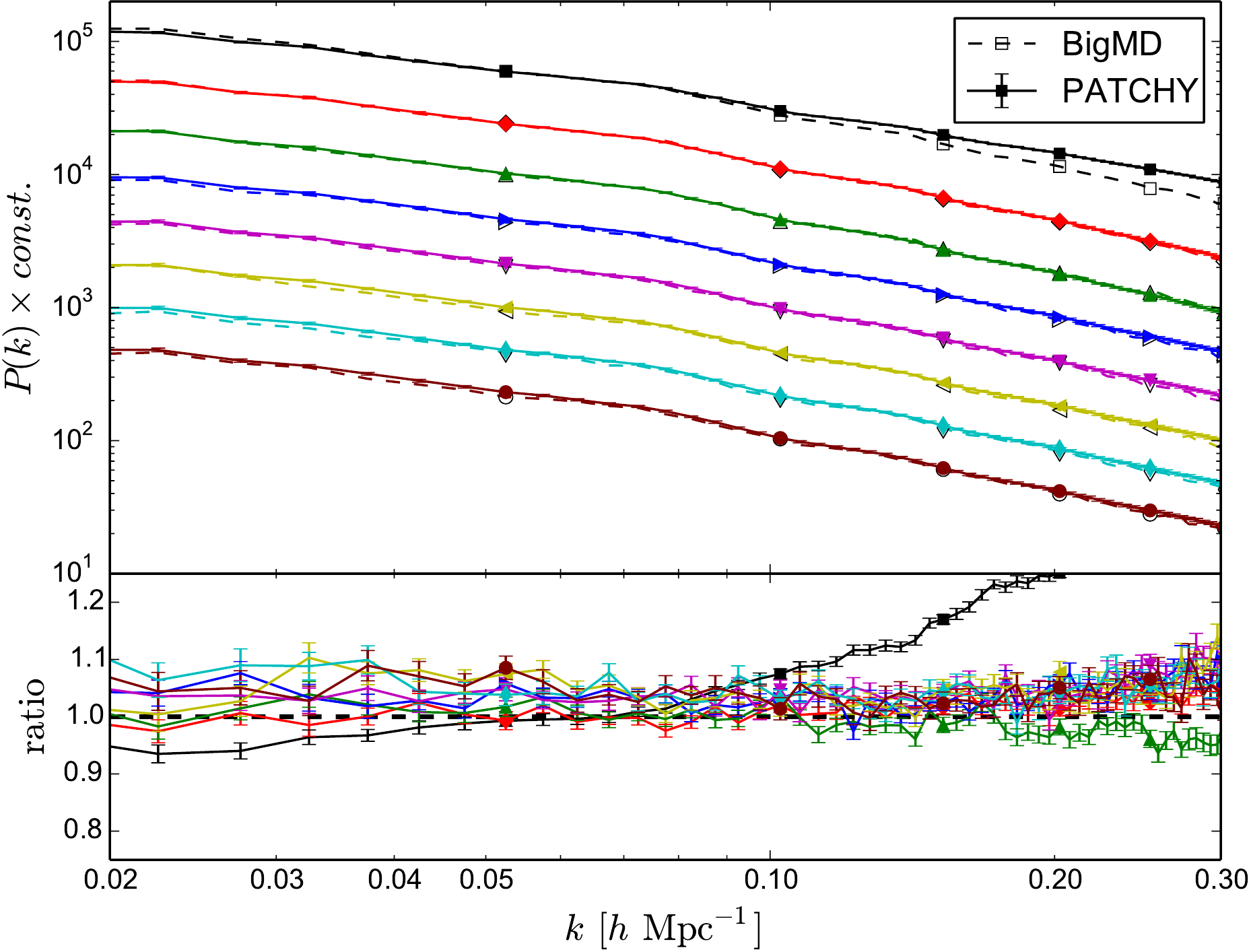}
\includegraphics[width=.47\textwidth]{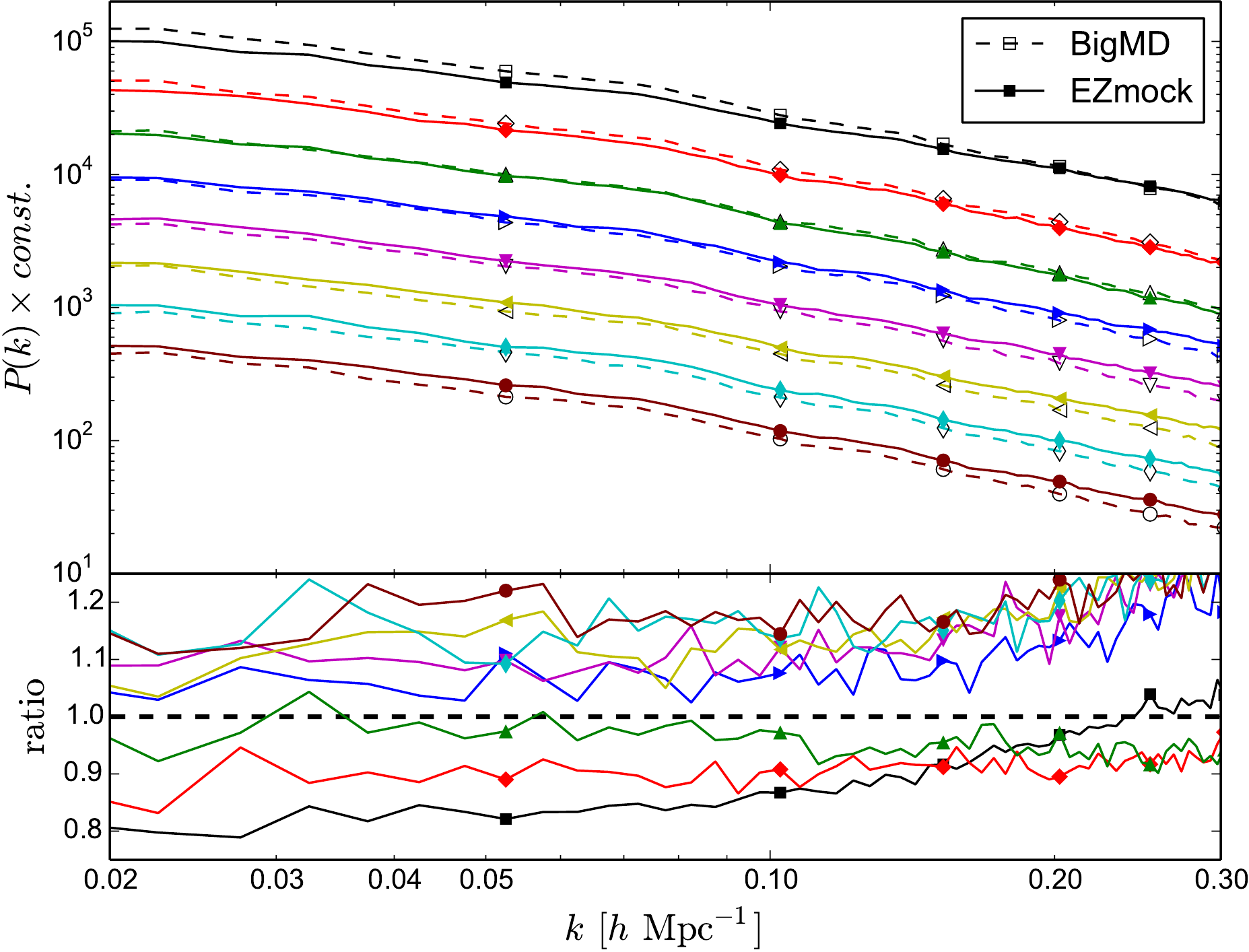}
\end{tabular}
\caption{Power spectra of  (left:) \textsc{patchy} and (right:) \textsc{EZmock} BDM mock catalogues in different \vmax{} bins after \vmax{} assignment.}
\label{fig:bdm_pt}
\end{figure*}

\subsection{Mass assignment in perturbation theory based mocks}
\label{sec:appPT}

We apply the \textsc{hadron}-code to  \textsc{patchy} and \textsc{EZmock} based catalogues which have been generated using initial conditions based on the ones corresponding to the Planck \textsc{BigMD} simulation but re-sampled to a lower resolution of $960^3$ cells. This reduces the cosmic variance, however, the  population of haloes based on the dark matter field has still a random component  \citep[cf.][]{patchyp1,ezmockp}.  
As in section \ref{sec:num}, we focus on the BDM halo catalogue from \textsc{BigMD} selected by \vmax{}. 

Fig.~\ref{fig:bdm_all} shows the power spectra of the entire \textsc{patchy} and \textsc{EZmock} mock halo catalogues compared to the BDM halo catalogue from \textsc{BigMD}, which serves as the reference for calibration of the mock catalogues. An analogous analysis has been conducted with FoF haloes and the results are shown in Appendix \ref{sec:pt}.
\textsc{EZmocks} uses a clouds-in-cell based population of haloes, whereas \textsc{patchy} uses counts-in-cell in the population step according to the negative binomial distribution modelling over-dispersion, i.e., the deviation from Poissonity \citep[][]{patchyp1}. This results in slightly better agreements in terms of the power spectra for \textsc{EZmocks} based catalogues when using  CIC estimators, as we do here and can be seen in Fig.~\ref{fig:bdm_all}. 
We assign \vmax{} to these two mock catalogues using the procedure described in \S \ref{sec:hadron}. However, as we have mentioned, we have still some degrees of freedom within this procedure, in  the definition of the cosmic web structures ($\lambda_{\rm th}$) and the \vmax{} threshold.
We calibrate these degrees of freedom according to the performance (power spectra of sub-samples) of the catalogues after the \vmax{} assignment. In this case, we employ  $\lambda_{\rm th}=-0.25$, and the \vmax{} threshold $V_{\rm th}=500\,$km\,s$^{-1}$. 
Moreover, we generate 100 \textsc{patchy} realizations with the same initial conditions, but  changing the random seeds  constructing the halo catalogues. This permits us to estimate the error bars for the 2- and 3-point statistics sharing the same large scale cosmic variance for a consistent comparison to the $N$-body simulation.
The comparison of power spectra for different \vmax{} bins is shown on the left and right panels of Fig.~\ref{fig:bdm_pt}  for \textsc{patchy} and \textsc{EZmock} mock catalogues, respectively.

\begin{figure}
\centering
\includegraphics[width=.47\textwidth]{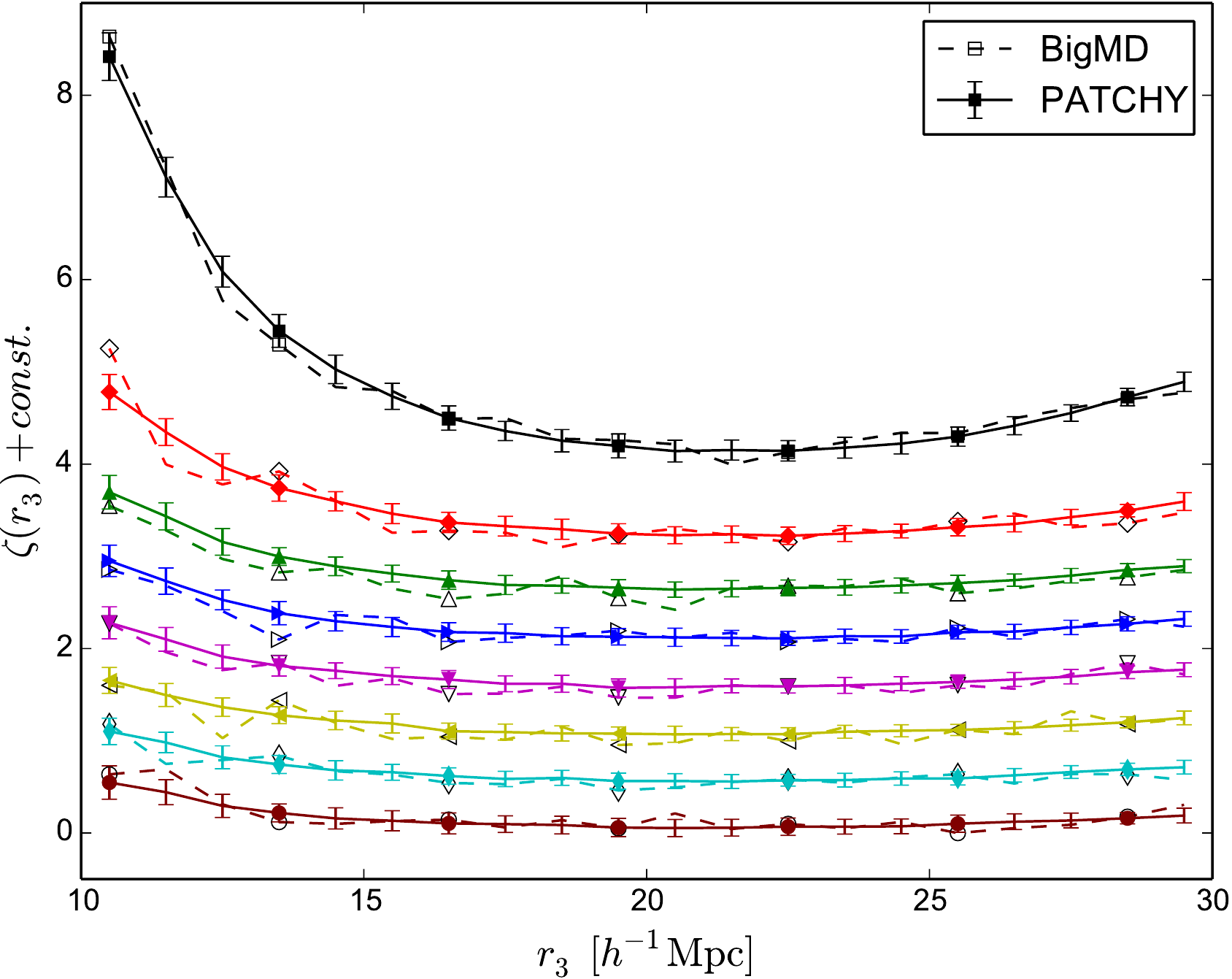}
\caption{3-point correlation functions of \textsc{patchy} BDM mock catalogues in different \vmax{} bins after \vmax{} assignment.}
\label{fig:3pcf_patchy}
\end{figure}

We find that \textsc{patchy} reaches a significantly higher accuracy in the biased tracers, due to the more precise dark matter field  used within the approach (ALPT vs Zel'dovich). In particular, the precision reached with \textsc{patchy} equals the one reached with the mass re-assignment tests using the $N$-body simulation (see \S \ref{sec:vmax3}), whereas the systematic deviations with \textsc{EZmock} grow to 20\% within $k<0.2\,h$\,Mpc$^{-1}$.
This means that the remarkable performance of \textsc{EZmock} in terms of the global 2- and 3-point statistics (on large scales) as shown in several works \citep[][]{ezmockp,ChuangComp2014} suffers from its crude dark matter density approximation (on small scales) with the mass assignment scheme presented in this work.
Let us therefore continue our analysis focused on \textsc{patchy}.
 Fig.~\ref{fig:3pcf_patchy} shows the great performance of the 2- and 3-point correlation functions  for different \vmax{} bins using \textsc{patchy}.
We see that the deviation found in the power spectrum for the most massive bin is also apparent on small scales in the 2-point correlation function (see black lines in Fig.~\ref{fig:3pcf_patchy}). Nevertheless, the BAO peak is matched within 1-$\sigma$ for all mass bins, showing an excellent agreement down to the smallest scales ($\sim5\,h^{-1}$Mpc), but for the most massive bin.
 The 3-point correlation function is essentially compatible with the $N$-body simulation for the different mass bins.
It is interesting to observe how the anisotropy increases towards higher masses, showing that the tracers with larger mass are less homogeneously distributed across the cosmic web.

\begin{figure}
\centering
\includegraphics[width=.47\textwidth]{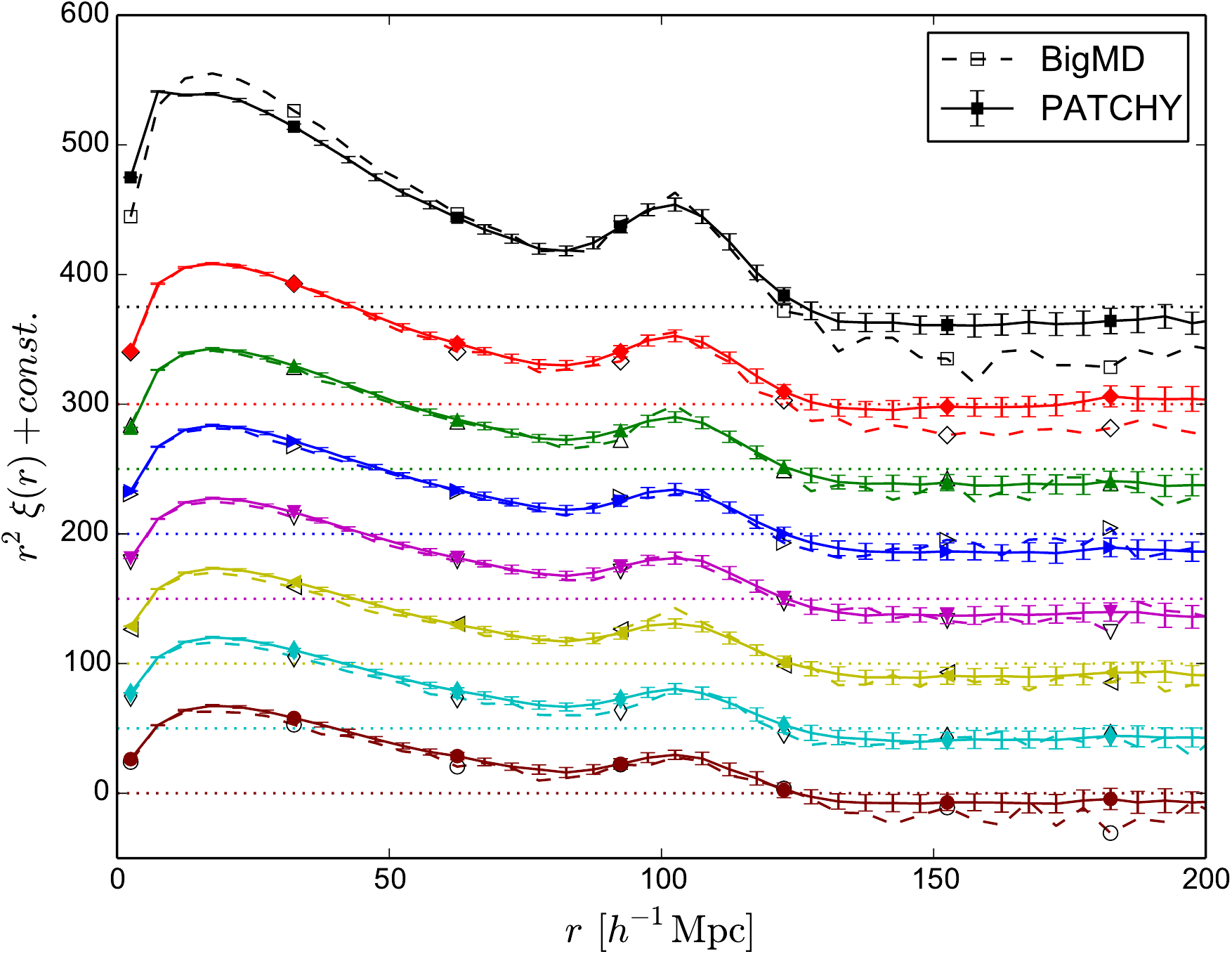}
\caption{2-point correlation functions of \textsc{patchy} BDM mock catalogues in different \vmax{} bins after \vmax{} assignment.}
\label{fig:xi_patchy}
\end{figure}

\section{Conclusions}
\label{sec:con}

In this work, we have studied the probabilistic dependence of the halo mass distribution as a function of local and non-local indicators, such as the local density, the cosmic web environment, and the halo-exclusion effect. 
We have found complex non-linear relations between the halo mass and the local density field, showing a degeneracy between parent haloes and subhaloes in certain density environments, as expected. Furthermore, we have used the non-local cosmic web environment information according to the eigenvalues of the tidal field tensor.  This permits us to find accurate statistical relations between the halo mass and the density and cosmic web environment. Such relations can be used to assign masses to a distribution of haloes. We dubbed the implementation of our method the \textsc{hadron} (Halo mAss Distribution ReconstructiON) code. We first have tested this on the halo distribution of the Planck BigMultiDark simulation by ignoring the actual information of their masses to reconstruct them using the statistical relations found in this work. We furthermore tested this method on a halo distribution produced by perturbation theory based codes, such as \textsc{patchy} and \textsc{EZmocks}. Our results show that accurate perturbation theory models are required to properly model the halo mass to density relation. In particular, augmented Lagrangian perturbation theory (ALPT), as opposed to the Zel'dovich approximation, permits us to dramatically reduce the errors.
We find that the resulting populations (classified into different mass bins) of haloes using ALPT agree in terms of power spectra within 1$\sigma$ up to scales of $k=0.2$ for different mass cuts, demonstrating that we recover the correct scale dependent bias on those scales. Only the most massive haloes (\vmax{}$\gsim550$ km\,s$^{-1}$) show a larger deviation. For these, we  find evidence of the halo-exclusion effect, as a clear improvement is achieved when assigning those high masses with a  minimum separation.
Furthermore, we have computed the two- and three-point correlation functions finding an excellent agreement for arbitrary mass cuts. 

This method can be applied for efficient massive production of mock halo or galaxy catalogues. Our work represents a quantitative application of the cosmic web classification. It can have further interesting applications in the multi-tracer analysis of the large-scale structure for future galaxy surveys.

\section*{acknowledgments}
CZ and CT acknowledge support from Tsinghua University, and 973 program No. 2013CB834906. CZ also thanks the support from MultiDark summer student program to visit the Instituto de F\'{i}sica Te\'{o}rica, (UAM/CSIC), Spain.
CC and FP were supported by the Spanish MICINNs Consolider-Ingenio 2010 Programme under grant MultiDark CSD2009-00064 and AYA2010-21231-C02-01 grant, the Comunidad de Madrid under grant HEPHACOS S2009/ESP-1473, and Spanish MINECOs Centro de Excelencia Severo Ochoa Programme under grant
SEV-2012-0249. 
GY acknowledges support from the Spanish MINECO under research grants
AYA2012-31101, FPA2012-34694 and Consolider Ingenio SyeC CSD2007-0050.

The BigMultiDark simulation suite have been performed in the SuperMUC  supercomputer at the Leibniz-Rechenzentrum in Munich,  thanks to the CPU time
 awarded by PRACE (proposal number 2012060963).
We also acknowledge PRACE for awarding us access to Curie
supercomputer based in France (project PA2259).
Some other computations were performed on HYDRA, the HPC-cluster of the
IFT-UAM/CSIC.

\bibliography{hadron}

\appendix

\section{FoF halo catalogues selected by mass}
\label{sec:fof}

This appendix shows the analysis performed using FoF haloes in an analogous way to the BDM halo analysis shown in the main text (see \S \ref{sec:num} and \ref{sec:appPT}). 

\subsection{$N$-body simulation based catalogues} 
\label{sec:nbody}

\begin{figure}
\centering
\includegraphics[width=.45\textwidth]{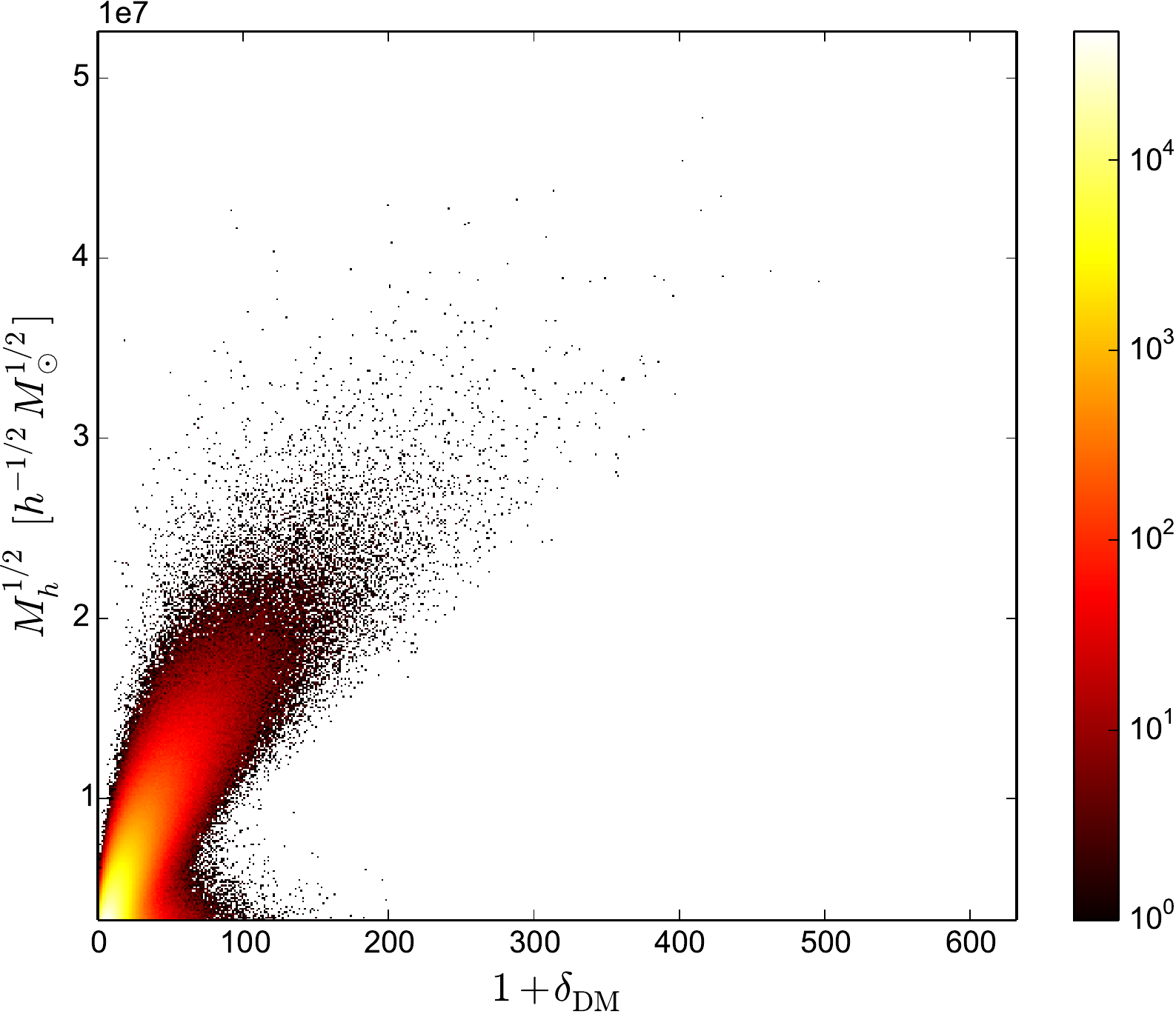}
\caption{Number of FoF haloes in \textsc{BigMD} with certain mass and local DM density.}
\label{fig:md}
\end{figure}

When applying the \textsc{hadron} method to FoF catalogues, we have to adopt mass assignment instead of \vmax{} assignment, since \vmax{} is not available for FoF haloes. In this case, the extracted mass--density relation is shown in Fig.~\ref{fig:md}.

\begin{figure}
\centering
\includegraphics[width=.47\textwidth]{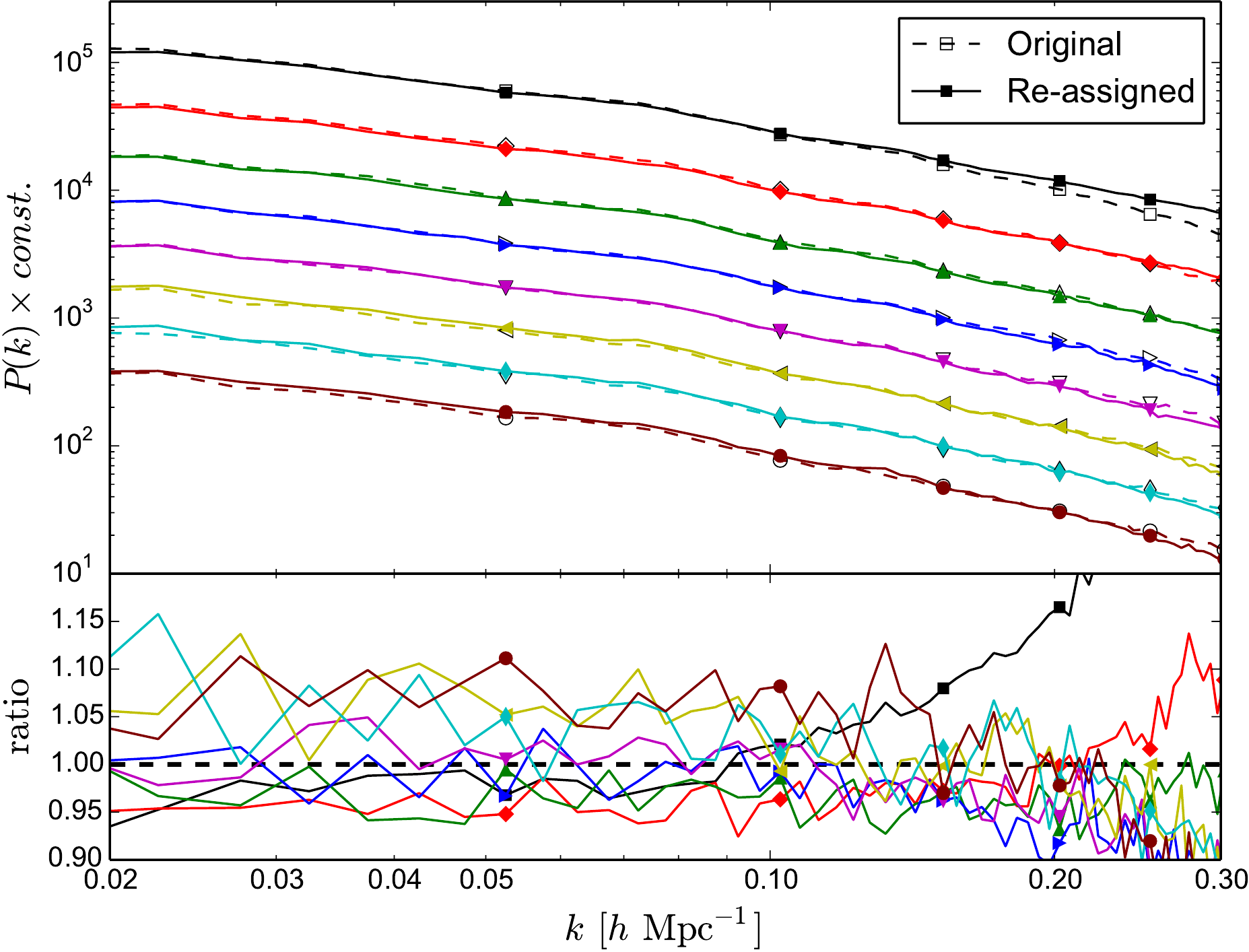}
\caption{Power spectra of re-assigned sub-catalogues with different halo masses, in comparison to the original FoF mock catalogues.}
\label{fig:pkmass}
\end{figure}

\begin{figure}
\centering
\includegraphics[width=.47\textwidth]{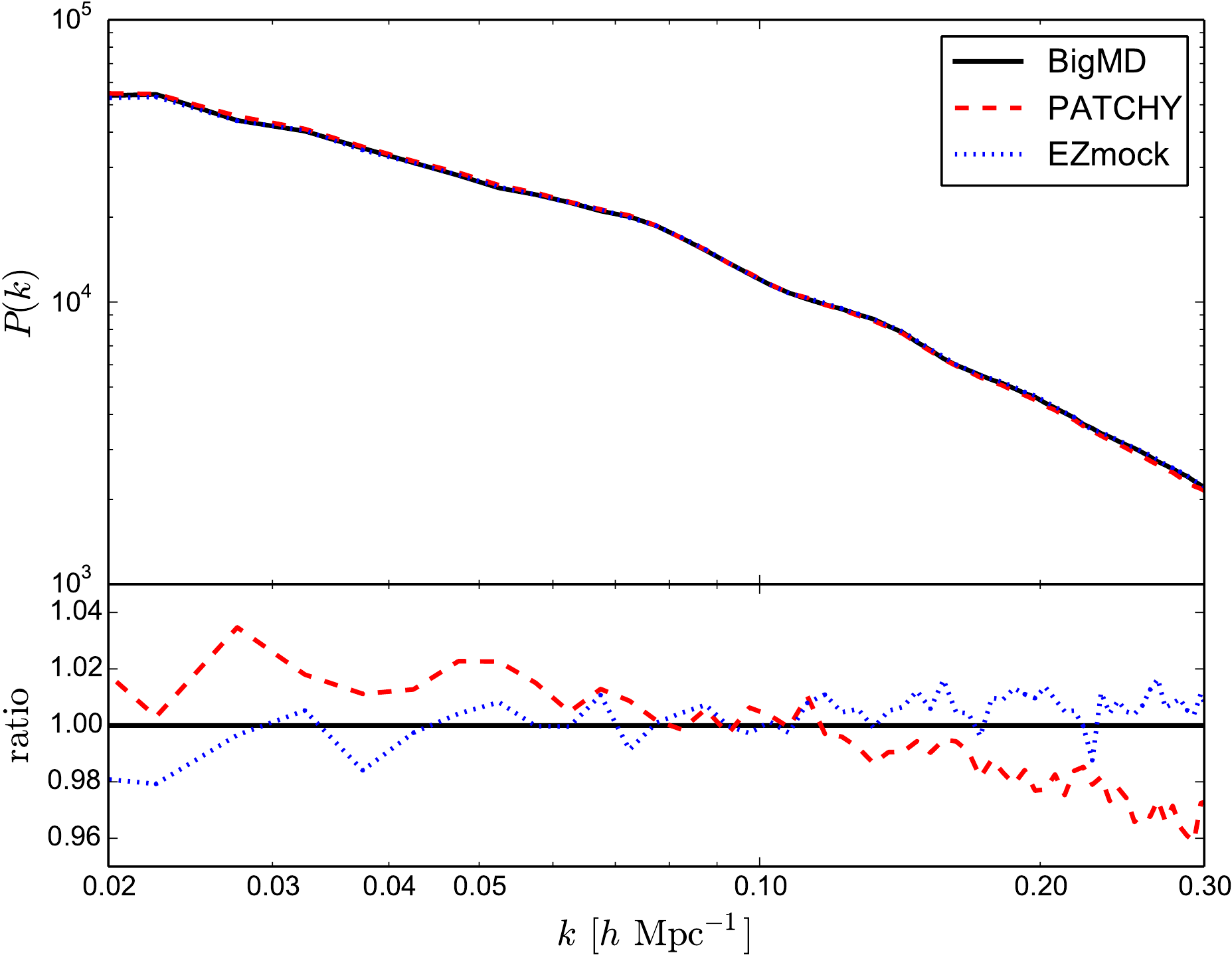}
\caption{Comparison of power spectra for the catalogues before mass assignment.}
\label{fig:fof_all}
\end{figure}

\begin{figure*}
\begin{tabular}{cc}
\includegraphics[width=.47\textwidth]{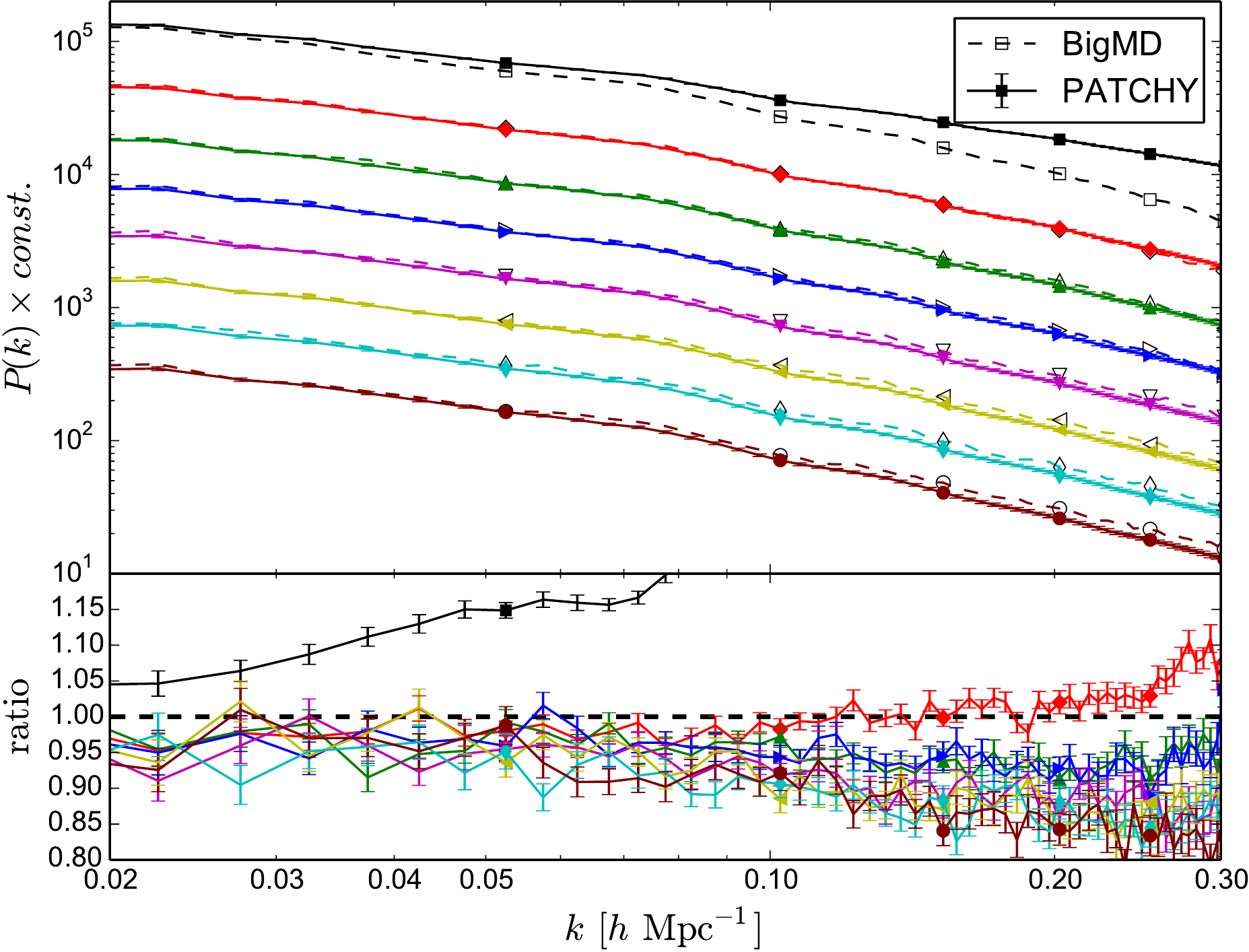}
\includegraphics[width=.47\textwidth]{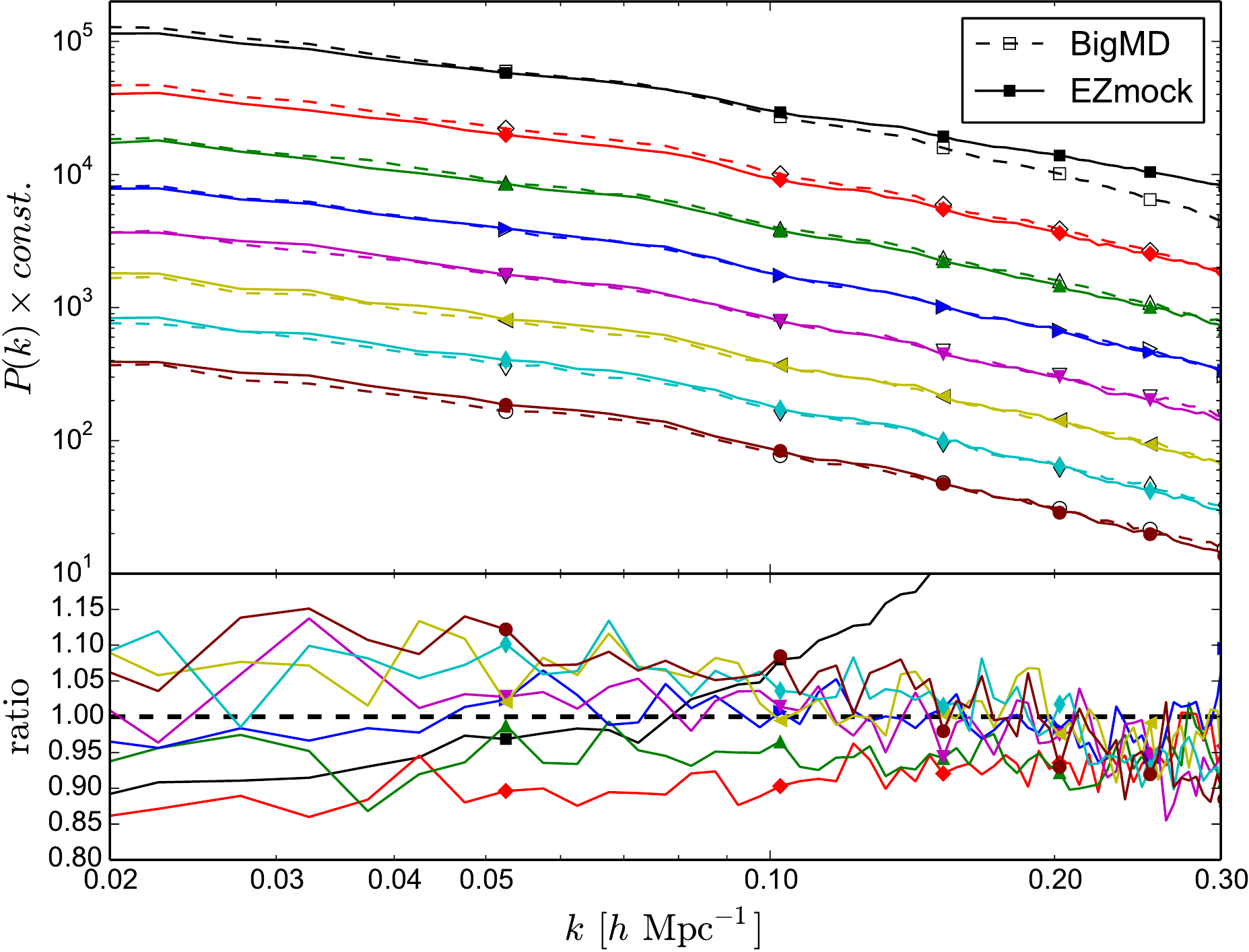}
\end{tabular}
\caption{Power spectra of (left:) \textsc{patchy} and (right:) \textsc{EZmock} FoF mock catalogues in different mass bins after mass assignment.}
\label{fig:fof_pt}
\end{figure*}

We then follow the same steps as for the \vmax{} assignment (see \S \ref{sec:num}), i.e. classify the cosmic web structures and employ a mass threshold. To justify the re-assigned result, we also divide the catalogues into 8 sub-samples of different halo masses, i.e. \{[$<$1.17), [1.17-1.33), [1.33-1.54), [1.54-1.84), [1.84-2.28), [2.28-3.08), [3.08-4.92), [$\geq$4.92]\}\,$\times 10^{13}\,h^{-1}M_\odot$. Fig.~\ref{fig:pkmass} displays the power spectra of the sub-samples drawn from the original \textsc{BigMD} FoF halo catalogue and that after mass re-assignment. The mass threshold for the exclusion operation are $2.51 \times 10^{13}\,h^{-1}M_\odot$ for both catalogues.

We note that the performance for FoF  is worse than for the BDM catalogue, reaching deviations of about 10\% already at $k\sim0.15\,h$\,Mpc$^{-1}$.

\subsection{Perturbation theory based catalogues}
\label{sec:pt}

For the FoF samples, the performance of \textsc{patchy} and \textsc{EZmock} is shown in Fig.~\ref{fig:fof_all}. With $\lambda_{\rm th}=-0.25$, and mass threshold of $2.51 \times 10^{13}\,h^{-1}M_\odot$ for the mass assignment procedure, the power spectra for different mass bins is shown on the left and right panels of  Fig.~\ref{fig:fof_pt} for \textsc{patchy} and \textsc{EZmock} mock catalogues, respectively.
The performance for FoF with  \textsc{patchy} is worse than for the BDM catalogues. We have seen in \ref{sec:nbody}, that this is also true for the $N$-body simulation. Nevertheless, another reason is that \textsc{patchy} has been designed to model over-dispersion, which is present in the BDM catalogues when taking the full population of haloes, including subhaloes, but not true for the FoF catalogue.

\end{document}